\documentclass[floats,aps,epsf,amsfonts]{revtex4}
\usepackage{enumerate}
\usepackage[dvips]{graphicx}

\newcommand{\calT}{{\cal T}}
\newcommand{\ttau}{\tilde\tau}

\newcommand{\Omegar}{\Omega_r}
\newcommand{\hOmegaf}{\hat\Omega_\varphi}
\newcommand{\hOmegar}{\hat\Omega_r}
\newcommand{\Omegaf}{\Omega_\varphi}
\newcommand{\U}{\left\langle U\right \rangle}
\newcommand{\hatU} {\hat{\left\langle U\right \rangle}_{\ttau}}
\newcommand{\hatUbg}{\hat{\left\langle U\right \rangle}_{\ttau 0}}

\newcommand{\pI}{\mathsf{p}}
\newcommand{\eI}{\mathsf{e}}

\begin{document}

\title{Beyond the geodesic approximation: conservative effects of the gravitational self-force in eccentric orbits 
around a Schwarzschild black hole}
\author{Leor Barack$^1$ and Norichika Sago$^2$}
\affiliation{
$^1$School of Mathematics, University of Southampton, Southampton
SO17 1BJ, United Kingdom, \\
$^2$Yukawa Institute for Theoretical Physics, Kyoto University,
Kyoto 606-8502, Japan}

\date{\today}

\begin{abstract}
We study conservative finite-mass corrections to the motion of a particle in a bound (eccentric) strong-field orbit around a Schwarzschild black hole. We assume the particle's mass $\mu$ is much smaller than the black hole mass $M$, and explore post-geodesic corrections of $O(\mu/M)$. Our analysis uses numerical data from a recently developed code that outputs the Lorenz-gauge gravitational self-force (GSF) acting on the particle along the eccentric geodesic. First, we calculate the $O(\mu/M)$ conservative correction to the periastron advance of the orbit, as a function of the (gauge-dependent) semilatus rectum and eccentricity. A gauge-invariant description of the GSF precession effect is made possible in the circular-orbit limit, where we express the correction to the periastron advance as a function of the invariant azimuthal frequency. We compare this relation with results from fully nonlinear numerical-relativistic simulations. In order to obtain a gauge-invariant measure of the GSF effect for fully eccentric orbits, we introduce a suitable generalization of Detweiler's circular-orbit ``redshift'' invariant. We compute the $O(\mu/M)$ conservative correction to this invariant, expressed as a function of the two invariant frequencies that parametrize the orbit. Our results are in good agreement with results from post-Newtonian calculations in the weak-field regime, as we shall report elsewhere. The results of our study can inform the development of analytical models for the dynamics of strongly gravitating binaries. They also provide an accurate benchmark for future numerical-relativistic simulations.
\end{abstract}

\maketitle

\section{Introduction and summary}

We recently reported \cite{Barack:2010tm} a numerical code for computing the gravitational self-force (GSF) experienced by a mass particle set in motion around a Schwarzschild black hole. The code takes as input the two parameters of an eccentric geodesic ($p$ and $e$, representing relativistic generalizations of the Keplerian semilatus rectum and eccentricity, respectively) and returns the various components of the GSF, evaluated in the Lorenz gauge, as functions along the geodesic orbit. Our current code does not take into account the back reaction from the GSF on the orbital motion; describing the radiative evolution of the orbit remains an important long-term goal of the self-force program \cite{Gair:2010iv}, strongly motivated within the context of experimental gravitational-wave physics \cite{Barack:2003fp}. However, GSF data from our current code (and from GSF codes by others \cite{Detweiler:2008ft,Keidl:2010pm,Akcay:2010dx}, so far restricted to circular orbits) already give access to some interesting new physics. Specifically, they allow a quantitative description of some conservative post-geodesic effects associated with the finiteness of the particle's mass. This description is {\em exact} at linear order in the particle's mass $\mu$ (within a controlled numerical error, which in practice can be kept very small).

Several such post-geodesic effects have already been analyzed. In \cite{Detweiler:2008ft} Detweiler calculated the conservative $O(\mu)$ correction to the value of the gauge-invariant ``redshift'' parameter ($dt/d\tau$, with $t$ and $\tau$  being the particle's coordinate and proper times, respectively; see below) along a circular orbit. In \cite{Barack:2009ey} we derived the conservative $O(\mu)$ shift in the frequency of the innermost stable circular orbit (ISCO). Most recently, in \cite{Barack:2010ny} (with Damour) we computed the $O(\mu)$ precession effect of the conservative GSF at the circular-orbit limit. These results were all successfully tested against the analytic predictions of post-Newtonian (PN) theory in the weak-field regime \cite{Detweiler:2008ft,Damour:2009sm,Barack:2010ny}, and, moreover, used to constrain some of the yet-unknown high-order PN parameters \cite{Blanchet:2009sd,Damour:2009sm,Blanchet:2010zd,Barack:2010ny}.

Indeed, GSF calculations already provide a valuable source of strong-field calibration data for various analytic models of the dynamics in binary systems. Huerta and Gair \cite{Huerta:2008gb} used circular-orbit GSF data from Ref.\ \cite{Barack:2007tm} to assess the influence of conservative GSF corrections on the long-term phase evolution in astrophysical extreme-mass-ratio inspirals. Lousto {\it et al.}~\cite{Lousto:2009mf,Lousto:2009ka} used the ISCO-shift result to inform an ``empirical'' formula (based also on results from fully nonlinear numerical simulations and on PN information) for the remnant masses and spins in binary black hole mergers. The value of the ISCO shift was also utilized by Favata \cite{Favata:2010yd} as an exact reference point in an exhaustive study of the relative performances of various PN methods. And, in a recent collaboration with Damour \cite{Barack:2010ny} we used GSF data for slightly eccentric orbits to constrain, for the first time, the strong-field shape of some of the analytic functions appearing in the effective one body (EOB) description of the conservative binary dynamics. In all these examples GSF data are valuable because they give a handle on an hitherto inaccessible, ``remote'' region of the binary parameter space, namely the domain of small mass-ratios and small separations. 

So far, analysis of the conservative effects of the GSF has been restricted to circular orbits or to orbits with an infinitesimally small eccentricity. Here we take a wide step forward, and study the conservative effects of the GSF for orbits with {\em large} eccentricities (``large'' in the sense of ``not necessarily small''---in practice our code can handle eccentricities up to about $0.7$; see below). In doing so we exploit the full capacity of our eccentric-orbit code for the first time. 

Eccentric geodesics of the Schwarzschild geometry are characterized by two frequencies: an azimuthal frequency $\Omega_\varphi$ and a radial frequency $\Omega_r$, associated with the periodic motions in the Schwarzschild coordinates $\varphi$ and $r$, respectively (the two frequencies will be defined more precisely below). Unlike in the analogous Keplerian problem where $\Omega_\varphi=\Omega_r$, here one always finds $\Omega_\varphi>\Omega_r$, giving rise to what we interpret as {\it periastron advance}. The conservative piece of the GSF induces $O(\mu)$ shifts in the values of $\Omega_\varphi$ and $\Omega_r$ and hence in the value of the periastron advance. Our first goal in this paper is to compute this GSF-induced correction to the periastron advance, denoted $\Delta\delta$, and we do so in Sec.\ \ref{Sec:advance}. We express  $\Delta\delta$ as a function of (the GSF-perturbed values of) $p$ and $e$---a convenient parametrization of the orbit which is commonly adopted in perturbative studies (e.g., \cite{Cutler:1994pb,Gair:2005ih,Sundararajan:2008zm}). Unfortunately, although the shift $\Delta\delta$ itself is gauge invariant (in a sense that will be made precise in Sec.\ \ref{subsec:gauge}), the parameters $p$ and $e$ are {\em not} invariant, so our numerical results for $\Delta\delta(p,e)$ must be interpreted with caution: they are valid specifically in the Lorenz gauge. A sample of our results for $\Delta\delta(p,e)$ is presented in Tables \ref{table:delta_data1} and \ref{table:delta_data2} and plotted in Fig.\ \ref{fig:delta}. Notably, we find that the correction $\Delta\delta(p,e)$ is {\it negative} for any $p,e$, i.e., the Lorenz-gauge GSF always acts to {\em reduce} the rate of periastron advance. 

The above calculation accounts only for the effect of the {\em conservative} piece of the GSF, and it assumes that its dissipative piece has been ``turned off''. Of course, in the physical problem the frequencies $\Omega_\varphi$ and  $\Omega_r$ evolve dissipatively, and over one radial period they drift by an amount of $O(\mu)$, which may well be comparable in magnitude to the conservative shift in these frequencies. One should therefore ask whether the conservative shift $\Delta\delta$ has any physical relevance (notwithstanding the gauge issue mentioned above). We will suggest an interpretation of our results that makes them meaningful even in the presence of dissipation: The periastron advance in the physical, dissipatively evolving system may still be defined through the accumulated $\varphi$-phase between two successive periastron passages (minima of the orbital radius $r$); we will observe that this ``true'' periastron advance is given through $O(\mu)$ by the {\em conservative} advance $\Delta\delta$ calculated for a certain ``average'' geodesic. This observation makes it possible, in principle, to incorporate the $O(\mu)$ precession effect in an approximate model of the orbital evolution (such as the one introduced recently by Gair {\it et al.}~in Ref.\ \cite{Gair:2010iv}) based on our results for $\Delta\delta$.

It is clearly desirable to have at hand a gauge-invariant measure of the $O(\mu)$ conservative effect---to allow, for example, a meaningful comparison with results from PN theory. What is required, more precisely, is a gauge-invariant {\it relation}, that expresses a nontrivial gauge-invariant quantity encoding the $O(\mu)$ effect as a function of two other gauge-invariant quantities parametrizing the perturbed orbit. A natural gauge-invariant parametrization of the orbit is provided by the pair $\{\Omega_\varphi,\Omega_r\}$\footnote{An important caveat is discussed in the last paragraph of Sec.\ \ref{subsec:prelim} and in Appendix \ref{App:sing}.}, but it is not immediately obvious what one might choose for a third nontrivial invariant. In Sec.\ \ref{Sec:U} we will propose such a third invariant, which is constructed from the GSF (and the regularized metric perturbation) along the orbit, and encodes the $O(\mu)$ conservative effect in a nontrivial way. Our gauge-invariant quantity, denoted $\U_{\tau}$, represents a natural generalization of Detweiler's ``redshift'' invariant: it is defined as the $\tau$-average of $dt/d\tau$ over a radial period, which is also equal to the ratio between the radial period measured in time $t$ and the radial period measured in time $\tau$. We then compute the conservative GSF correction $\Delta\U_{\tau}$ as a function of the (GSF-corrected) frequencies $\Omega_\varphi$ and $\Omega_r$. Our numerical results for the gauge-invariant relation $\Delta\U_{\tau}(\Omega_\varphi,\Omega_r)$ are displayed in Table \ref{table:U_data}. Preliminary comparison of these results with PN expressions in the weak-field regime shows a very good agreement. A detailed comparison with PN theory will be presented elsewhere \cite{inprep}.

In the above description we have (for simplicity) left unmentioned a certain subtlety related to our choice of a time coordinate, and we come back to this point now. It is known \cite{Barack:2005nr,Damour:2009sm,Barack:2010ny} that the monopole piece of the Lorenz-gauge metric perturbation has a peculiar behavior at $r\to\infty$, which, however, can be cured using a simple ``normalization'' of the Schwarzschild time coordinate, in the form $t\to \hat t=(1+\alpha) t$, where $\alpha$ is a certain $O(\mu)$ constant which depends on the particle's orbit [see Eq.\ (\ref{alpha}) below]. The time $\hat t$ (unlike $t$) reflects the asymptotic flatness of the perturbed spacetime, and at $r\to\infty$ it coincides with the natural time coordinate used in PN studies. Since one of our main goals here is to provide useful data for PN comparisons, we choose to work with the time coordinate $\hat t$ throughout our analysis. In particular, we define the orbital frequencies with respect to time $\hat t$, not $t$ (and call them $\hat\Omega_{\varphi}$ and $\hat\Omega_r$ to reflect this), and we similarly define $\U_{\tau}$ through $d\hat t/d\tau$, not $dt/d\tau$. We note, however, that the quantity $\Delta\delta$ is insensitive to the time coordinate chosen (as it depends only on the ratio $\Omega_{\varphi}/\Omega_r=\hat\Omega_{\varphi}/\hat\Omega_r$).

In Sec.\ \ref{Sec:circ} we return to consider the periastron advance, this time focusing on the circular-orbit limit ($e\to 0$). This limit defines a one-parameter family of orbits, each of which nonetheless characterized by {\em two} invariant frequencies $\hat\Omega_{\varphi}$ and $\hat\Omega_r$ (the latter associated with an infinitesimal $e$-perturbation of the circular orbit). This allows for a gauge-invariant description of the GSF precession effect in this case, e.g., by expressing $\delta$ as a function $\hat\Omega_{\varphi}$. In previous work \cite{Barack:2010ny} we have already analyzed the precession effect of the GSF for slightly eccentric orbits. Here we reproduce our results in a form that allows us a direct comparison with recent results from fully nonlinear numerical relativistic (NR) simulations by Mrou\'{e} {\it et al.}~\cite{Mroue:2010re}. We illustrate the feasibility and potential benefit of such comparisons in order to motivate more detailed study. 

To help readers navigate through this work we give, in Table \ref{table:symbols}, a key to our main symbols and essential notation. The table can serve as a quick reference guide for readers wishing to use our numerical results without delving into the technical details of their derivation. Throughout this work we use metric signature ${-}{+}{+}{+}$ and ``geometrized'' units with $c=G=1$ (with the mass of the central black hole providing a natural unit for both time and distance). Physical units can be restored by multiplying all distances by $G/c^2$, all frequencies by $c^3/G$, etc.

\begin{table}[htb] 
\begin{tabular}{l|l|l}
\hline\hline
symbol & meaning & where defined \\
\hline\hline
$\mu$, $M$ & particle's mass, black hole's mass   & -- \\
$q$ & $\mu/M$, small mass ratio  & -- \\
$t,r,\theta,\varphi$   & Schwarzschild/Lorenz-gauge coordinates   & -- \\
$F^{\rm cons}_{\alpha}$ & conservative piece of the GSF ($\propto\mu^2$)&  -- \\
$X_0$  &  value of a quantity $X$ at the geodesic limit ($\mu\to 0$)  & -- \\
$\Delta X$  &  the $O(\mu)$ conservative GSF correction to $X$  & e.g., Eqs.\ (\ref{def:deltaSF}), (\ref{DeltaU}) \\
$p$, $e$ & semilatus rectum (per $M$) and eccentricity & Eq.\ (\ref{tr1tr2}) \\
$\chi$    & orbital radial phase (``true anomaly'')    & Eq.\ (\ref{tr}) \\
$E$, $L$ & particle's energy and angular momentum per $\mu$  & Eq.\ (\ref{tEtL})   \\
 $\Omega_r$, $\Omega_{\varphi}$ & radial and azimuthal orbital frequencies & Eqs.\ (\ref{Omegar}), (\ref{Omegaf}) \\
$T$  & radial period (in time $t$) & --   \\ 
$\cal T$  & radial period (in proper time $\tau$) & --  \\ 
$\delta$  & fractional periastron advance (dissipation ignored) & Eq.\ (\ref{tdelta})   \\ 
$\delta_{\rm true}$  & fractional periastron advance (dissipation included) & Eq.\ (\ref{delta_true})   \\ 
$\hat t$  & normalized (``asymptotically flat'') time coordinate            & Eq.\ (\ref{hatt}) \\
$\hat X$ &  a quantity $X$ redefined with respect to time $\hat t$  & e.g., Eq.\ (\ref{homega})\\
$\U_{\tau}$ &  generalized redshift invariant &  Eq.\ (\ref{def:U}) \\
$r_\circ$ & radius ($r$ value) of a circular orbit & -- \\
\hline\hline
\end{tabular}
\caption{Essential notation and key to main symbols.}
\label{table:symbols}
\end{table}

\section{Precession effect of the GSF}\label{Sec:advance}

\subsection{Preliminaries: geodesic motion}\label{subsec:prelim}

Let us begin by recalling a few results from the theory of geodesic motion in Schwarzschild spacetime. Throughout our presentation we use a subscript `0' to distinguish ``geodesic'' quantities from their GSF-perturbed counterparts; thus, for example, $u_0^{\alpha}$ and $u^{\alpha}$ will denote the tangent four-velocity associated with the geodesic and GSF-perturbed orbits, respectively. Timelike geodesics in Schwarzschild geometry constitute a 2-parameter family, parametrized by two constants of motion: the specific energy $E_0\equiv -u_{t0}$ and specific angular momentum $L_0\equiv u_{\varphi0}$. Here $u_{\alpha0}=g_{\alpha\beta}u_0^{\beta}$, where $g_{\alpha\beta}$ is the background Schwarzschild metric with mass $M$, and we have assumed (without loss of generality) that the geodesics lie in the ``equatorial plane'', $\theta=\pi/2$ (hence $u_0^{\theta}=0$). We will be interested in the subset of {\em bound} geodesics, i.e., ones confined to a radii range $r_0^{-}\leq r\leq r_0^{+}$ for some $r_0^{-}>4M$ (``periastron'') and $r_0^{+}\geq r_0^{-}$ (``apastron'') (there are no bound timelike geodesics with $r_0^{-}\leq 4M$). The pair $(r_0^{-},r_0^{+})$ can be used as an alternative parametrization of these geodesics, which we call {\em eccentric}. The special (one-parameter subset of) eccentric geodesics with $r_0^{-}=r_0^{+}$ are called {\em circular}. 

For eccentric geodesics it is convenient to define the (adimensionalized) semilatus rectum $p_0$ and eccentricity $e_0$ through
\begin{equation}\label{p0e0}
r_0^{-}=\frac{p_0M}{1+e_0}, \quad\quad r_0^{+}=\frac{p_0M}{1-e_0},
\end{equation}
and reparametrize the orbits using the pair $(p_0,e_0)$. The structure of the $(p_0,e_0)$ parameter space is described, e.g., in Sec.\ II.A of \cite{Barack:2010tm} (see in particular Fig.\ 1 therein). Each eccentric geodesic has a unique value of $(p_0,e_0)$ within the range $0\leq e_0<1$ with $p_0>6+2e_0$. Geodesics located along the {\em separatrix} $p_0=6+2e_0$ are marginally unstable (and will not remain bound under small perturbations of $e_0$ and/or $p_0$). Stable circular orbits have $e_0=0$ and $p_0>6$. The special geodesic with $(p_0,e_0)=(6,0)$ (corresponding in the $e_0$--$p_0$ plane to the intersection of the separatrix with the $e_0=0$ axis) is the ISCO. 

In terms of $p_0$ and $e_0$, the geodesic specific energy and angular momentum are given by
\begin{equation}\label{E0L0}
E_0=\left[\frac{(p_0-2-2e_0)(p_0-2+2e_0)}{p_0(p_0-3-e_0^2)}\right]^{1/2},
\quad\quad
L_0=\frac{p_0M}{(p_0-3-e_0^2)^{1/2}}.
\end{equation}
The orbital radius varies in time according to the formula
\begin{equation}\label{rofchi}
r=r_0(\chi)=\frac{p_0M}{1+e_0\cos\chi},
\end{equation}
where the radial phase $\chi$ (``true anomaly'') is related to the coordinate time $t=t_0(\chi)$ through 
\begin{equation}\label{dtdchi}
\frac{dt_0}{d\chi} =
\frac{Mp_0^2}{(p_0-2-2e_0\cos\chi)(1+e_0\cos\chi)^2}
\sqrt{\frac{(p_0-2-2e_0)(p_0-2+2e_0)}{p_0-6-2e_0\cos\chi}},
\end{equation}
with initial condition $t_0(\chi=0)=t^-$ where $t^-$ corresponds to a periastron passage [i.e., $r_0(\chi(t^-))=r_0^{-}$].
The azimuthal phase of the orbit, $\varphi=\varphi_0(\chi)$, can be obtained as a function of $\chi$ by solving 
\begin{equation}\label{dphidchi}
\frac{d\varphi_0}{d\chi} =
\sqrt{\frac{p_0}{p_0-6-2e_0\cos\chi}}
\end{equation}
with initial condition $\varphi(\chi=0)=\varphi^-$, where $\varphi^-$ is the $\varphi$-phase at $t=t^-$.
We note that both $t_0(\chi)$ and $\varphi_0(\chi)$ are monotonically increasing functions.   

The $t$-period of the radial motion (i.e., the $t$-time interval between two successive periastron passages) and the frequency associated with it are given by 
\begin{equation}\label{Omegar0}
T_0 = \int_{0}^{2\pi}\frac{dt_0}{d\chi}\,d\chi,
\quad \quad \Omega_{r0}= 2\pi/T_0,
\end{equation}
where the integrand can be expressed as a function of $\chi$ using Eq.\ (\ref{dtdchi}). The azimuthal phase accumulated over one radial period $T_0$, i.e., $\Phi_0\equiv\varphi_0(\chi=2\pi)-\varphi_0(\chi=0)$, is given by 
\begin{equation}\label{Phi0}
\Phi_0 =\int_{0}^{2\pi}\frac{d\varphi_0}{d\chi}\,d\chi
=4\left(\frac{p_0}{p_0-6-2e_0}\right)^{1/2}{\rm ellipK}\left(\frac{-4e_0}{p_0-6-2e_0}\right),
\end{equation}
where we have substituted from Eq.\ (\ref{dphidchi}) and where 
$
{\rm ellipK}(\gamma)\equiv \int_{0}^{\pi/2}(1-\gamma\sin^2x)^{-1/2}dx
$
is the complete elliptic integral of the first kind. 
The {\it azimuthal} frequency $\Omega_{\varphi0}$ is defined as the $t$-average of $(d\varphi/dt)_0$ over a radial period, which may also be expressed as 
\begin{equation}\label{Omegaphi0}
\Omega_{\varphi0} = \Phi_0/T_0= \Omega_{r0}\times[\Phi_0/(2\pi)].
\end{equation}
It can be shown that, for any given $0\leq e_0<1$, $\Phi_0$ is a monotonically decreasing function of $p_0$, with $\Phi_0\to 2\pi$ as $p_0\to\infty$. Hence $\Phi_0>2\pi$ for all eccentric orbits, with the excess $\Phi_0-2\pi$ representing {\em periastron advance}. We shall denote the average periastron advance per radian by $\delta_0$:
\begin{equation}\label{delta0}
\delta_0\equiv \Phi_0/(2\pi)-1>0.
\end{equation}
From Eq.\ (\ref{Omegaphi0}) we then have the relation 
\begin{equation}
\Omega_{\varphi0}/\Omega_{r0}=1+\delta_0.
\end{equation}

It is instructive to examine some asymptotic characteristics of $\delta_0$.
For a fixed eccentricity we have the large-$p_0$ expansion 
\begin{equation}\label{delta_largep}
\delta(p_0\gg 1)=3p_0^{-1}+\frac{1}{4}(54+3e_0^2)p_0^{-2}+O(p_0^{-3}),
\end{equation}
and we note that $e_0$ only enters $\delta_0$ at sub-leading order---and even there the effect of eccentricity is suppressed due to the smallness of the term $3e_0^2$ compared with $54$.
We also observe that $\delta_0$ does not vanish at the circular-orbit limit; rather, we find
\begin{equation}
\delta(e_0\ll 1)=\sqrt{p_0/(p_0-6)}-1+O(e_0^2).
\end{equation}
Note that at the ISCO (where $\Omega_{r0}=0$) $\delta_0$ diverges as $\sim (p_0-6)^{-1/2}$.
Lastly, consider the asymptotic behavior near the separatrix, $p_0-6-2e_0\ll 1$ (for $e_0>0$):
\begin{equation}\label{delta_separatrix}
\delta_0 \simeq \frac{1}{\pi} \left(\frac{3+e_0}{2e_0}\right)^{1/2}\ln\left(\frac{64e_0}{p_0-6-2e_0}\right).
\end{equation}
Note the logarithmic divergence at the separatrix limit. We see that the form of divergence of $\delta_0$ at the ISCO limit $(p_0,e_0)\to(6,0)$ depends on the direction (in the $p_0$--$e_0$ plane) from which this limit is taken, illustrating the singular nature of the ISCO point.

It should be noted that the pair of frequencies $(\Omega_{r0},\Omega_{\varphi0})$ does {\em not} constitute a one-to-one parametrization of eccentric geodesic orbits in Schwarzschild geometry. The transformation $(p_0,e_0)\to (\Omega_{r0},\Omega_{\varphi0})$ [and also $(E_0,L_0)\to (\Omega_{r0},\Omega_{\varphi0})$] becomes singular along a certain curve in the parameter space, well outside the separatrix (see Fig.\ \ref{fig:sing} in Appendix \ref{App:sing}); one finds that for each orbit lying on the inner side of the singular curve there exists a ``dual'' orbit on the outer side, which is physically distinct but has precisely the same frequencies $(\Omega_{r0},\Omega_{\varphi0})$. This little-known property of the parameter space (we were unable to find any reference to it in the literature) 
is further discussed in Appendix \ref{App:sing}.

\subsection{The GSF-perturbed orbit}

Now consider a pointlike particle of mass $\mu\ll M$ moving in the Schwarzschild geometry. At the limit $\mu\to 0$ the particle's orbit is a geodesic of the background geometry, and we assume here this geodesic belongs to the class of eccentric geodesics discussed above. When the finiteness of $\mu$ is taken into account, we say that the particle experiences a GSF ($\propto\mu^2$), and the motion is no longer strictly a geodesic of the background geometry. The GSF has a dissipative effect, which removes energy and angular momentum from the system (via gravitational radiation) and gives rise to a gradual inspiral. The GSF also has a conservative piece, which affects, for example, the precession rate of the orbit. The splitting of the GSF into dissipative and conservative pieces is formally defined in terms of the retarded and advanced metric perturbations (see, e.g., \cite{Hinderer:2008dm}); in Ref.\ \cite{Barack:2010tm} 
we describe how each of the two pieces can be constructed from the full GSF in practice, taking advantage of the particular symmetries of Schwarzschild geodesics. We will proceed here by making the (nonphysical) assumption that the dissipative piece of the GSF has been ``turned off'', and that the particle is moving under the influence of the conservative piece of the GSF alone, denoted $F_{\alpha}^{\rm cons}$. We shall come back to discuss the dissipative effect in the next subsection.

We shall assume that the orbit remains strictly bound under the small perturbation caused by $F^{\alpha}_{\rm cons}$ (this is only allowed because we are ignoring dissipation), and denote the values of the new, ``perturbed'' radial turning points by $r^{-}$ and $r^{+}$. (We note that the values of $r^{-},r^{+}$ are gauge dependent, as are the values of many other perturbed quantities we define below. We will describe the construction of gauge-invariant quantities in the next section; in the meantime, for concreteness, we may assume that all perturbed quantities are given in a particular gauge---e.g., the Lorenz gauge.) We then define $p$ and $e$ via
\begin{equation}\label{tr1tr2}
r^{-}=\frac{p M}{1+e}, \quad\quad r^{+}=\frac{p M}{1-e},
\end{equation}
and use the pair $(p,e)$ to parametrize the perturbed orbit. We also define 
\begin{equation}\label{tEtL}
E\equiv -u_t, \quad\quad  L\equiv u_\varphi,
\end{equation}
where $u_{\alpha}=g_{\alpha\beta}u^{\beta}$ with $u^{\beta}$ being the 4-velocity tangent to the perturbed orbit, taken to be normalized with respect to the {\em background} metric: $g_{\alpha\beta}u^{\alpha}u^{\beta}=-1$. Note that $E$ and $L$ are not conserved along the orbit, and do not have the interpretation of energy and angular momentum. 
From symmetry, the orbit remains equatorial even under the effect of the GSF: $u^{\theta}=0$. 

We further assume that the perturbed orbit remains periodic in time, with radial period $T$ interpreted as the perturbed value of $T_0$ (that such a periodic solution to the perturbed equations of motion exists will be confirmed below by explicit construction). We then choose a radial phase parameter $\chi$ so defined that the radius along the perturbed orbit varies according to 
\begin{equation}\label{tr}
r=r(\chi)=\frac{p M}{1+e\cos\chi},
\end{equation}
in analogy with the geodesic formula (\ref{rofchi}) (we use the same symbol $\chi$ as in the geodesic case for mere notational brevity; this should not lead to confusion). The
effect of the GSF on the shape of the orbit is then encoded in the relations $t=t(\chi)$ and $\varphi=\varphi(\chi)$ along the perturbed orbit, which depend explicitly on the GSF (see below). The perturbed period and radial frequency are obtained through 
\begin{equation}\label{Omegar}
T = \int_{0}^{2\pi}\frac{dt}{d\chi}\,d\chi,
\quad\quad \Omegar= 2\pi/T,
\end{equation}
and the perturbed accumulated phase $\Phi$ and azimuthal frequency are given by 
\begin{equation}\label{Omegaf}
\Phi =\int_{0}^{2\pi}\frac{d\varphi}{d\chi}\,d\chi,
\quad\quad \Omegaf = \Phi/T.
\end{equation}
Finally, the conservative GSF-perturbed value of the periastron advance reads
\begin{equation}\label{tdelta}
\delta=\Phi/(2\pi)-1=\Omegaf/\Omegar-1.
\end{equation}

In what follows we will obtain an explicit expression for $\delta$ as a function of $p,e$ (which will, of course, involve $F_{\alpha}^{\rm cons}$). We will then define the $O(\mu)$ ``post-geodesic'' conservative correction to $\delta$ as
\begin{equation}\label{def:deltaSF}
\Delta\delta \equiv \delta(p,e)-\delta_0(p,e),
\end{equation}
where $\delta_0(p,e)$ is the ``background'' value obtained from Eq.\ (\ref{delta0}) with Eq.\ (\ref{Phi0}), replacing $(p_0,e_0)\to(p,e)$. Notice that our definition of the ``$O(\mu)$ correction'' employs only the {\em perturbed} parameters $(p,e)$. We are not considering here the alternative definition $\tilde\Delta\delta \equiv \delta(p,e)-\delta_0(p_0,e_0)$, which, of course, differs from $\Delta\delta$ at the leading order, $O(\mu)$. The motivation for considering the correction $\Delta$ rather than $\tilde\Delta$ will become clear in the next section, where we reparametrize the orbit using two gauge-invariant quantities---the frequencies $\Omega_r$ and $\Omega_\varphi$---and calculate the $\Delta$ correction to a third invariant, expressed as a function of the two perturbed frequencies. This $\Delta$ correction will then be truly gauge invariant, unlike the corresponding $\tilde\Delta$ correction, which would depend on the background frequencies $\Omega_{r0}$ and $\Omega_{\varphi0}$. We remind, however, that, in our present discussion, neither $\tilde\Delta\delta(p,e;p_0,e_0)$ {\em nor}  $\Delta\delta(p,e)$ constitute gauge-invariant relations. This is because the orbital parameters $p,e$ themselves are gauge-ambiguous. In this section we will be calculating $\Delta\delta$ specifically in the Lorenz gauge.

\subsection{Conservative GSF correction to the periastron advance} \label{subsec:GSFcorrection}

Recalling Eq.\ (\ref{Omegaf}), let us write  
\begin{equation}\label{tildeomega}
\frac{d\varphi}{d\chi}\equiv \varphi_{\chi}(\chi;p,e)  = \varphi_{\chi0}(\chi;p,e)+\Delta\varphi_{\chi}(\chi;p,e),
\end{equation}
where the ``background'' quantity $\varphi_{\chi0}(\chi;p,e)$ is obtained from Eq.\ (\ref{dphidchi}) replacing $(p_0,e_0)\to (p,e)$. From Eq.\ (\ref{tdelta}) we then have
\begin{equation}
\delta=\left[\frac{1}{2\pi}\int_{0}^{2\pi}\varphi_{\chi0}(\chi;p,e)d\chi-1\right]+
\frac{1}{2\pi}\int_{0}^{2\pi}\Delta\varphi_{\chi}(\chi;p,e)d\chi,
\end{equation}
and, identifying the term in square brackets as $\delta_0(p,e)$, we obtain from the definition in (\ref{def:deltaSF})
\begin{equation}\label{Deltadelta1}
\Delta\delta(p,e)= \frac{1}{2\pi}\int_{0}^{2\pi}\Delta\varphi_{\chi}(\chi;p,e)d\chi.
\end{equation}

To obtain an expression for $\Delta\varphi_{\chi}(\chi;p,e)$, let us start by writing 
\begin{equation}\label{varphichi}
\varphi_{\chi}=\frac{\dot{\varphi}}{\dot{r}} \frac{dr}{d\chi}
=\frac{e L(\chi)|\sin\chi|}{p M[E^2(\chi)-V(r(\chi),L(\chi))]^{1/2}},
\end{equation}
where hereafter an overdot denotes $d/d\tau$, with $\tau$ being proper time along the perturbed geodesic.
Here we have used $\dot{\varphi}=g^{\varphi\varphi}u_{\varphi}=r^{-2}L$, substituting for $r(\chi)$ from Eq.\ (\ref{tr}); the factor $dr/d\chi$ was evaluated using Eq.\ (\ref{tr}) again; and to evaluate $\dot{r}$ we used the normalization of the perturbed 4-velocity, $g_{\alpha\beta}u^{\alpha} u^{\beta}=-1$, giving $\dot{r}^2=E^2-V(r(\chi),L(\chi))$, with the effective potential 
\begin{equation}
V(r,L)=\left(1-\frac{2M}{r}\right)\left(1+\frac{L^2}{r^2}\right).
\end{equation}
The ``background'' quantity $\varphi_{\chi0}(\chi;p,e)$ in Eq.\ (\ref{tildeomega}) can be obtained from Eq.\ (\ref{varphichi}) by fixing $p,e,\chi$ [hence also fixing $r(\chi)$] and replacing $E(\chi;p,e)\to E_0(p,e)$ and $L(\chi;p,e)\to L_0(p,e)$ [where $E_0(p,e)$ and $L_0(p,e)$ are the geodesic functional relations obtained from Eq.\ (\ref{E0L0}) with $(p_0,e_0)\to (p,e)$]. The GSF correction $\Delta\varphi_{\chi}=\varphi_{\chi}-\varphi_{\chi0}$ can therefore be obtained at $O(\mu)$ by considering the $\Delta$-variation of $\varphi_\chi$ in Eq.\ (\ref{varphichi}) with respect to $E$ and $L$:
\begin{equation}\label{Deltaomega}
\Delta\varphi_{\chi}=
\left.\frac{\partial\varphi_{\chi}}{\partial E}\right|_{E_0,L_0} \Delta E
+
\left.\frac{\partial\varphi_{\chi}}{\partial L}\right|_{E_0,L_0} \Delta L.
\end{equation}
Here the partial derivatives are taken with fixed $p,e,\chi$ (hence also fixed $r$) and evaluated at $(E_0(p,e),L_0(p,e))$, and  we have introduced $\Delta E(\chi;p,e)\equiv E(\chi;p,e)-E_0(p,e)$ and $\Delta L(\chi;p,e)\equiv L(\chi;p,e)-L_0(p,e)$. With Eqs.\ (\ref{Deltadelta1}) and (\ref{Deltaomega}), the task of calculating $\Delta\delta$ thus reduces to obtaining the $O(\mu)$ corrections $\Delta E$ and $\Delta L$.

To obtain $\Delta E$ and $\Delta L$ we need to consider the $t$ and $\varphi$ components of the particle's equation of motion. These read, respectively, 
$\mu\dot{u}_{t}=-F^{\rm cons}_t$ and $\mu\dot{u}_{\varphi}=F^{\rm cons}_\varphi$, or, with the definitions of Eq.\ (\ref{tEtL}),
\begin{equation}\label{def:deltaSFpe}
\mu\dot{E}=-F^{\rm cons}_t,
\quad\quad
\mu\dot{L}=F^{\rm cons}_\varphi.
\end{equation}
It is useful to think of $F^{\rm cons}_t$ and $F^{\rm cons}_\varphi$ as (periodic) functions of $\chi$ along the perturbed geodesic. We define the two $O(\mu)$ functions 
\begin{equation}\label{calEL}
{\cal E}(\chi)\equiv -\mu^{-1}\int_0^{\chi}F_t^{\rm cons}(\chi')\frac{d\tau}{d\chi'}d\chi',
\quad\quad
{\cal L}(\chi)\equiv \mu^{-1}\int_0^{\chi}F_\varphi^{\rm cons}(\chi')\frac{d\tau}{d\chi'}d\chi'.
\end{equation}
The factor $d\tau/d\chi$ in the integrands need only be evaluated at leading order in $\mu$ [since the expressions in Eq.\ (\ref{calEL}) are already $O(\mu)$], and we can therefore use for this purpose the geodesic expressions given earlier. Writing $d\tau/d\chi=(d\tau/dt)_0(dt_0/d\chi)=(1-2M/r_0)E_0^{-1}(dt_0/d\chi)$ and then substituting from Eqs.\ (\ref{E0L0}), (\ref{rofchi}) and (\ref{dtdchi}), we find, at leading order in $\mu$,
\begin{equation}\label{dtaudchi}
\frac{d\tau}{d\chi}=\frac{M p^{3/2}}{(1+e\cos\chi)^2}\sqrt{\frac{p-3-e^2}{p-6-2e\cos\chi}}.
\end{equation}
Note we are using here the perturbed parameters $(p,e)$ instead of $(p_0,e_0)$, which is allowed since interchanging $(p,e)\leftrightarrow(p_0,e_0)$ affects the expressions in Eq.\ (\ref{calEL}) only at $O(\mu^2)$. In Eq.\ (\ref{calEL}) it is sufficient, at our working order, to evaluate the GSF components along the background geodesic $(p_0,e_0)$. Given numerical data for the conservative GSF along the geodesic (the kind of data provided by our code \cite{Barack:2010tm}), the functions ${\cal E}(\chi)$ and ${\cal L}(\chi)$ can be evaluated numerically for any $\chi$ through $O(\mu)$. In fact, it is sufficient to evaluate these two functions for $0\leq \chi\leq \pi$, given the ``reflection'' symmetry ${\cal E}(\chi)={\cal E}(2\pi-\chi)$ and ${\cal L}(\chi)={\cal L}(2\pi-\chi)$ [see Ref.\ \cite{Barack:2010tm}, where we explain that the conservative $t,\varphi$ components of the GSF have the antisymmetry property $F_t^{\rm cons}(\chi)=-F_t^{\rm cons}(2\pi-\chi)$ and similarly for $F_\varphi^{\rm cons}$, and note that the factors $d\tau/d\chi$  in Eq.\ (\ref{calEL})  are reflection-symmetric].

Integrating Eqs.\ (\ref{def:deltaSFpe}), we now obtain 
\begin{eqnarray}\label{EL}
E(\chi)&=&E(0)+{\cal E}(\chi)=E_0+\Delta E(0)+{\cal E}(\chi),
\nonumber\\
L(\chi)&=&L(0)+{\cal L}(\chi)=L_0+\Delta L(0)+{\cal L}(\chi),
\end{eqnarray}
where $E(0)$ and $L(0)$ are the values of $E$ and $L$ at the periastron ($\chi=0$), and the $O(\mu)$ quantities $\Delta E(0)\equiv E(0;p,e)-E_0(p,e)$ and $\Delta L(0)\equiv L(0;p,e)-L_0(p,e)$ are the shifts in the values of $E$ and $L$ at the periastron due to the conservative GSF. 
We identify
\begin{eqnarray}\label{DeltaEDeltaL}
\Delta E(\chi)&=&\Delta E(0)+{\cal E}(\chi),
\nonumber\\
\Delta L(\chi)&=&\Delta L(0)+{\cal L}(\chi).
\end{eqnarray}

The values of the shifts $\Delta E(0)$ and $\Delta L(0)$ are found in the following manner. At the periastron ($r=r^{-}$, $\chi=0$) and apastron ($r=r^{+}$, $\chi=\pi$), where $\dot{r}=0$, it follows from the normalization of the perturbed 4-velocity that
\begin{equation}\label{norm}
E^2(0)=V(r^{-},L(0)), \quad\quad
E^2(\pi)=V(r^{+},L(\pi)).
\end{equation}
The $O(\mu)$ piece of these relations (holding $p,e$---hence also $r^{\pm}$---fixed) is given by 
\begin{eqnarray} \label{eq1}
E \Delta E(0)&=&\left(1-\frac{2M}{r^{-}}\right)\frac{L}{(r^{-})^2}\Delta L(0),
\\
E \Delta E(\pi)&=&\left(1-\frac{2M}{r^{+}}\right)\frac{L}{(r^{+})^2}\Delta L(\pi),
\end{eqnarray}
where $\Delta E(\pi)$ and $\Delta L(\pi)$ are the $O(\mu)$ shifts in the values of $E$
and $L$ at the apastron. These two are related to the shifts at the periastron through
\begin{eqnarray}
\Delta E(\pi)&=&\Delta E(0)+{\cal E}(\pi),
\\ \label{eq4}
\Delta L(\pi)&=&\Delta L(0)+{\cal L}(\pi).
\end{eqnarray}
Solving the four equations (\ref{eq1})--(\ref{eq4}) simultaneously for $\Delta E(0)$, $\Delta L(0)$, $\Delta E(\pi)$ and $\Delta L(\pi)$, with the substitutions $r^{-}=Mp/(1+e)$ and $r^{+}=Mp/(1-e)$, we obtain
\begin{equation}\label{deltaE0}
\Delta E(0)=\frac{(1+e)^2(p-2-2e)}{4e(p-3-e^2)}
\left[(1-e)^2(p-2+2e)B{\cal L}(\pi)-{\cal E}(\pi)\right],
\end{equation}
\begin{equation}\label{deltaL0}
\Delta L(0)=\frac{1}{4e(p-3-e^2)}
\left[(1-e)^2(p-2+2e){\cal L}(\pi)-{\cal E}(\pi)/B\right],
\end{equation}
with
\begin{equation}
B=\frac{L_0(p,e)}{E_0(p,e) M^2 p^3}=\frac{1}{M p^{3/2}[(p-2)^2-4e^2]^{1/2}},
\end{equation}
where we have substituted for $E_0(p,e)$ and $L_0(p,e)$ from Eq.\ (\ref{E0L0}).
With these expressions, the functions $\Delta E(\chi)$ and $\Delta L(\chi)$ of Eq.\ (\ref{DeltaEDeltaL}) are now fully specified given the parameters $p,e$ and the GSF functions ${\cal E}(\chi), {\cal L}(\chi)$.

Finally, substituting for $\Delta E$ and $\Delta L$ from Eq.\ (\ref{DeltaEDeltaL}) in Eq.\ (\ref{Deltaomega}), we
obtain the rather unwieldy but explicit result
\begin{eqnarray}\label{Deltaomega2}
\Delta\varphi_{\chi} &=&
\frac{p(p-3-e^2)^{1/2}[(p-2)^2-4e^2]^{1/2}}{e^2(p-6-2e\cos\chi)^{3/2}}
\left[\frac{{\cal E}(\pi)}{4\cos^2(\chi/2)}-\frac{{\cal E}(\chi)}{\sin^2\chi}
\right]
\nonumber\\
&&-\frac{p^{-1/2}(p-3-e^2)^{1/2}}{M e^2(p-6-2e\cos\chi)^{3/2}}
\left[\frac{(1-e)^2(p-2+2e){\cal L}(\pi)}{4\cos^2(\chi/2)}
-\frac{[p(1+e^2)-2(1+3e^2)+2e(p-3-e^2)\cos\chi]{\cal L}(\chi)}{\sin^2\chi}
\right].
\nonumber\\
\end{eqnarray}
Recall $\Delta\delta$ is obtained from $\Delta\varphi_{\chi}$ as prescribed in Eq.\ (\ref{Deltadelta1}). Notice in that equation that, since $\Delta\delta$ and $\Delta\varphi_{\chi}$ are of $O(\mu)$, we are allowed to replace $(p,e)\to (p_0,e_0)$ in their argument; this only results in an error of $O(\mu^2)$ which we neglect in our treatment. We also note that $\Delta\varphi_{\chi}$ has the ``reflection'' symmetry mentioned above, i.e., $\Delta\varphi_{\chi}(\chi)=\Delta\varphi_{\chi}(2\pi-\chi)$. This allows us to fold the integral piece $\int_{\pi}^{2\pi}$ over onto $\int_0^{\pi}$. With these modifications, we write our final result for $\Delta\delta$ in the form 
\begin{equation}\label{Deltadelta}
\Delta\delta(p_0,e_0)= \frac{1}{\pi}\int_{0}^{\pi}\Delta\varphi_{\chi}(\chi;p_0,e_0)d\chi,
\end{equation}
where $\Delta\varphi_{\chi}(\chi;p_0,e_0)$, given in Eq.\ (\ref{Deltaomega2}) with the replacement $(p,e)\to (p_0,e_0)$, only contains quantities which are evaluated along the background geodesic $(p_0,e_0)$.

Let us summarize the construction of $\Delta\delta$. One starts with (numerical) data sets for the conservative GSF components $F^{\rm cond}_t$ and 
$F^{\rm cond}_\varphi$, evaluated as functions of $\chi$ along the background geodesic $(p_0,e_0)$ between $\chi=0$ and $\chi=\pi$. These data are integrated via Eq.\ (\ref{calEL}) to yield the indefinite integrals ${\cal E}(\chi)$ and ${\cal L}(\chi)$, which in turn are used to construct the function $\Delta\varphi_{\chi}(\chi)$ via Eq.\ (\ref{Deltaomega2}). A second integration [Eq.\ (\ref{Deltadelta})] finally produces $\Delta\delta(p_0,e_0)$. Note in this procedure the GSF data are integrated over twice. It is possible to simplify this ``integral of integral'' structure by integrating by parts in Eq.\ (\ref{Deltadelta}). We find in practice, however, that this does not lead to significant simplification in the numerical implementation, and we hence prefer to leave our working formula for $\Delta\delta$ in its form (\ref{Deltadelta}).

Inspecting Eq.\ (\ref{Deltaomega2}), one may be worried about the apparent divergence of the terms $\propto \sin^{-2}\chi$ at $\chi=0,\pi$ and of the terms $\propto \cos^{-2}(\chi/2)$ at $\chi=\pi$. In fact, the function $\Delta\varphi_{\chi}$ is perfectly regular at these two points (and anywhere else within the domain $0\leq\chi\leq\pi$). To see this, it suffices to notice the following. First, at the limit $\chi\to 0$ we have ${\cal E,L}\propto\chi^2$, since the two functions are even in $\chi$ and they obviously vanish at $\chi\to 0$ [from their definition in Eq.\ (\ref{calEL}), recalling $F^{\rm cond}_t$ and $F^{\rm cond}_\varphi$ are regular functions of $\chi$ along the orbit]. Hence ${\cal E}/\sin^2\chi$ and ${\cal L}/\sin^2\chi$ have finite limits as $\chi\to 0$. Second, at the limit $\chi\to\pi$ we have the even Taylor expansion ${\cal E}={\cal E}(\pi)+\frac{1}{2}(\chi-\pi)^2{\cal E}''(\pi)+O(\chi-\pi)^4$ [recalling the reflection symmetry ${\cal E}(\chi)={\cal E}(\pi-\chi)$], and similarly for ${\cal L}$; substituting these expansions in Eq.\ (\ref{Deltaomega2}) reveals that the limit $\chi\to \pi$ of this expression is finite as well. Nonetheless, a direct numerical implementation of Eq.\ (\ref{Deltadelta}) does require some special care near $\chi=0,\pi$. In Appendix \ref{AppA} we describe our method for dealing with this technical subtlety in practice.

\subsection{Numerical results}

Our numerical algorithm for calculating the GSF is described in detail in Ref.\ \cite{Barack:2010tm}, where we also discuss the various sources of numerical error and our method for estimating the error bars on the numerical data.   
Our code returns the GSF as a function of $\chi$ along a given geodesic defined by the input parameters $p_0,e_0$. Our code can handle eccentricities in the range $0\leq e_0\lesssim 0.7$ and $p_0$ values between the separatrix and $\sim 100$; outside this range our current algorithm is less effective and the computation burden becomes prohibitive (see \cite{Barack:2010tm} for details). We should point out that even in the above ``workable'' part of the parameter space, there are narrow ``stripes'' in the $e_0$--$p_0$ plane which are currently inaccessible to our code. These inaccessible stripes correspond to orbits for which the frequency ratio $\Omega_{\varphi 0}/\Omega_{r0}$ is close to a small integer (mainly 2 or 3). For such ``resonant'' orbits, our algorithm for calculating the dipole piece of the metric perturbation (which is based on a frequency domain method) becomes ineffective. Hence, for example, the orbits with $(p_0,e_0)=(8,0.1)$ and $(8,0.2)$ have $\Omega_{\varphi 0}/\Omega_{r0}$ very close to 2, and are currently inaccessible to our code. We have discussed this problem in Ref.\ \cite{Barack:2010tm} and are hoping to address it in future work. In the current study, however, we will content ourselves with simply avoiding the problematic orbits.

\begin{table}[htb]
\begin{minipage}{2.2in}
\begin{tabular}{llll}
\hline\hline
$p_0$ & $e_0$ & $q^{-1}\Delta\delta$ & $q^{-1}\Delta\delta/\delta_0$ \\
\hline\hline
$6.1$ & $0.02$ & $-146(2)$ & $-20.7(2)$ \\
\hline
$6.2$ & $0.05$ & $-57.0(2)$ & $-11.71(5)$ \\
\hline
$6.3$ & $0.1$ & $-41.9(1)$ & $-10.23(3)$ \\
\hline
$6.4$ & $0.1$ & $-19.71(5)$ & $-6.12(2)$ \\
\hline
$6.5$ & $0.1$ & $-12.28(3)$ & $-4.51(1)$ \\
$6.5$ & $0.2$ & $-31.15(6)$ & $-9.36(2)$ \\
\hline
$6.7$ & $0.1$ & $-6.45(1)$ & $-3.010(6)$ \\
$6.7$ & $0.2$ & $-9.45(1)$ & $-4.062(6)$ \\
$6.7$ & $0.3$ & $-27.03(3)$ & $-9.30(1)$ \\
\hline
$7.0$ & $0.1$ & $-3.372(9)$ & $-2.024(6)$ \\
$7.0$ & $0.2$ & $-4.089(6)$ & $-2.359(3)$ \\
$7.0$ & $0.3$ & $-5.909(7)$ & $-3.157(4)$ \\
$7.0$ & $0.4$ & $-11.99(1)$ & $-5.516(5)$ \\
$7.0$ & $0.45$ & $-24.64(2)$ & $-9.764(8)$ \\
$7.0$ & $0.49$ & $-127.30(9)$ & $-37.25(3)$ \\
$7.0$ & $0.499$ & $-1279.4(9)$ & $-268.3(2)$ \\
$7.0$ & $0.4999$ & $-12735(7)$ & $-2075(1)$ \\
\hline\hline
\\
\\
\\
\end{tabular}
\end{minipage}
\begin{minipage}{2.2in}
\begin{tabular}{llll}
\hline\hline
$p_0$ & $e_0$ & $q^{-1}\Delta\delta$ & $q^{-1}\Delta\delta/\delta_0$ \\
\hline\hline
$7.5$ & $0.1$ & $-1.614(4)$ & $-1.298(3)$ \\
$7.5$ & $0.2$ & $-1.788(3)$ & $-1.411(2)$ \\
$7.5$ & $0.3$ & $-2.134(3)$ & $-1.629(2)$ \\
$7.5$ & $0.4$ & $-2.778(3)$ & $-2.015(2)$ \\
$7.5$ & $0.5$ & $-4.070(3)$ & $-2.733(2)$ \\
\hline
$8.0$ & $0.3$ & $-1.140(1)$ & $-1.101(1)$ \\
$8.0$ & $0.4$ & $-1.347(1)$ & $-1.264(1)$ \\
$8.0$ & $0.5$ & $-1.682(2)$ & $-1.516(1)$ \\
\hline
$8.5$ & $0.1$ & $-0.6156(5)$ & $-0.7276(6)$ \\
$8.5$ & $0.2$ & $-0.6497(9)$ & $-0.762(1)$ \\
$8.5$ & $0.3$ & $-0.7106(8)$ & $-0.822(1)$ \\
$8.5$ & $0.4$ & $-0.8052(9)$ & $-0.914(1)$ \\
$8.5$ & $0.5$ & $-0.9461(9)$ & $-1.046(1)$ \\
\hline
$9.0$ & $0.1$ & $-0.4299(4)$ & $-0.5861(5)$ \\
$9.0$ & $0.2$ & $-0.4493(6)$ & $-0.6089(8)$ \\
$9.0$ & $0.3$ & $-0.4832(5)$ & $-0.6483(7)$ \\
$9.0$ & $0.4$ & $-0.5346(6)$ & $-0.7070(7)$ \\
$9.0$ & $0.5$ & $-0.6082(6)$ & $-0.7894(8)$ \\
\hline\hline
\\
\\
\end{tabular}
\end{minipage}
\begin{minipage}{2.2in}
\begin{tabular}{llll}
\hline\hline
$p_0$ & $e_0$ & $q^{-1}\Delta\delta$ & $q^{-1}\Delta\delta/\delta_0$ \\
\hline\hline
$10$ & $0.1$ & $-0.2392(3)$ & $-0.4111(5)$ \\
$10$ & $0.2$ & $-0.2471(2)$ & $-0.4230(4)$ \\
$10$ & $0.3$ & $-0.2606(2)$ & $-0.4433(3)$ \\
$10$ & $0.4$ & $-0.2807(2)$ & $-0.4731(3)$ \\
$10$ & $0.5$ & $-0.3084(3)$ & $-0.5138(4)$ \\
\hline
$12$ & $0.1$ & $-0.10011(8)$ & $-0.2415(2)$ \\
$12$ & $0.2$ & $-0.102104(9)$ & $-0.24580(2)$ \\
$12$ & $0.3$ & $-0.105563(4)$ & $-0.25322(1)$ \\
$12$ & $0.4$ & $-0.110620(2)$ & $-0.264025(5)$ \\
$12$ & $0.5$ & $-0.11752(3)$ & $-0.27867(6)$ \\
\hline
$15$ & $0.1$ & $-0.03960(1)$  & $-0.13604(5)$ \\
$15$ & $0.2$ & $-0.03996(4)$ & $-0.1371(1)$ \\
$15$ & $0.3$ & $-0.04057(4)$ & $-0.1389(2)$ \\
$15$ & $0.4$ & $-0.04149(4)$ & $-0.1417(1)$ \\
$15$ & $0.5$ & $-0.04278(3)$ & $-0.1455(1)$ \\
\hline
$20$ & $0.1$ & $-0.01353(1)$ & $-0.06932(8)$ \\
$20$ & $0.2$ & $-0.01348(3)$ & $-0.0690(2)$ \\
$20$ & $0.3$ & $-0.01338(3)$ & $-0.0684(1)$ \\
$20$ & $0.4$ & $-0.01327(3)$ & $-0.0677(1)$ \\
$20$ & $0.5$ & $-0.01314(3)$ & $-0.0669(1)$ \\
\hline\hline
\end{tabular}
\end{minipage}
\caption{
Numerical results for the conservative (Lorenz-gauge) GSF correction to the periastron advance per radian, $\Delta\delta$, for a sample of $p_0,e_0$ values. $q\equiv \mu/M$ is the small mass ratio. The last column in each table shows the GSF correction as a fraction of the ``background'' advance $\delta_0$. Parenthetical figures indicate the estimated uncertainty in the last displayed decimals (due to numerical error); thus, e.g., $-19.71(5)$ stands for $-19.71\pm 0.05$. Note that the effect of the Lorenz-gauge GSF is to {\em reduce} the periastron advance ($\Delta\delta<0$). Note also, in the data for $p_0=7.0$, the manifest linear growth of $\Delta\delta$ with the inverse separatrix distance $(p_0-6-2e_0)^{-1}$; this behavior is discussed in the text.}
\label{table:delta_data1}
\end{table}

\begin{figure}[htb]
\includegraphics[width=8.9cm]{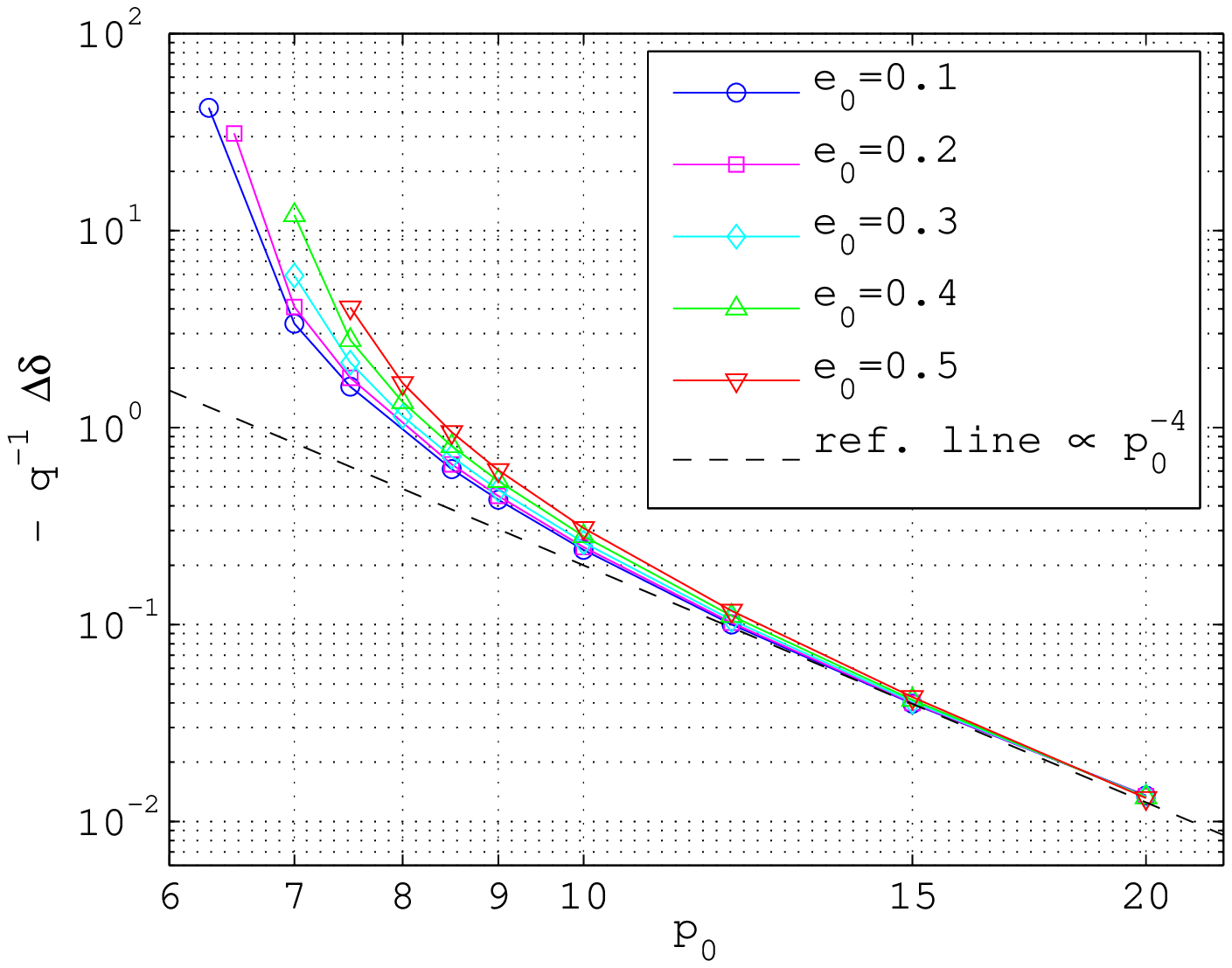}
\includegraphics[width=8.9cm]{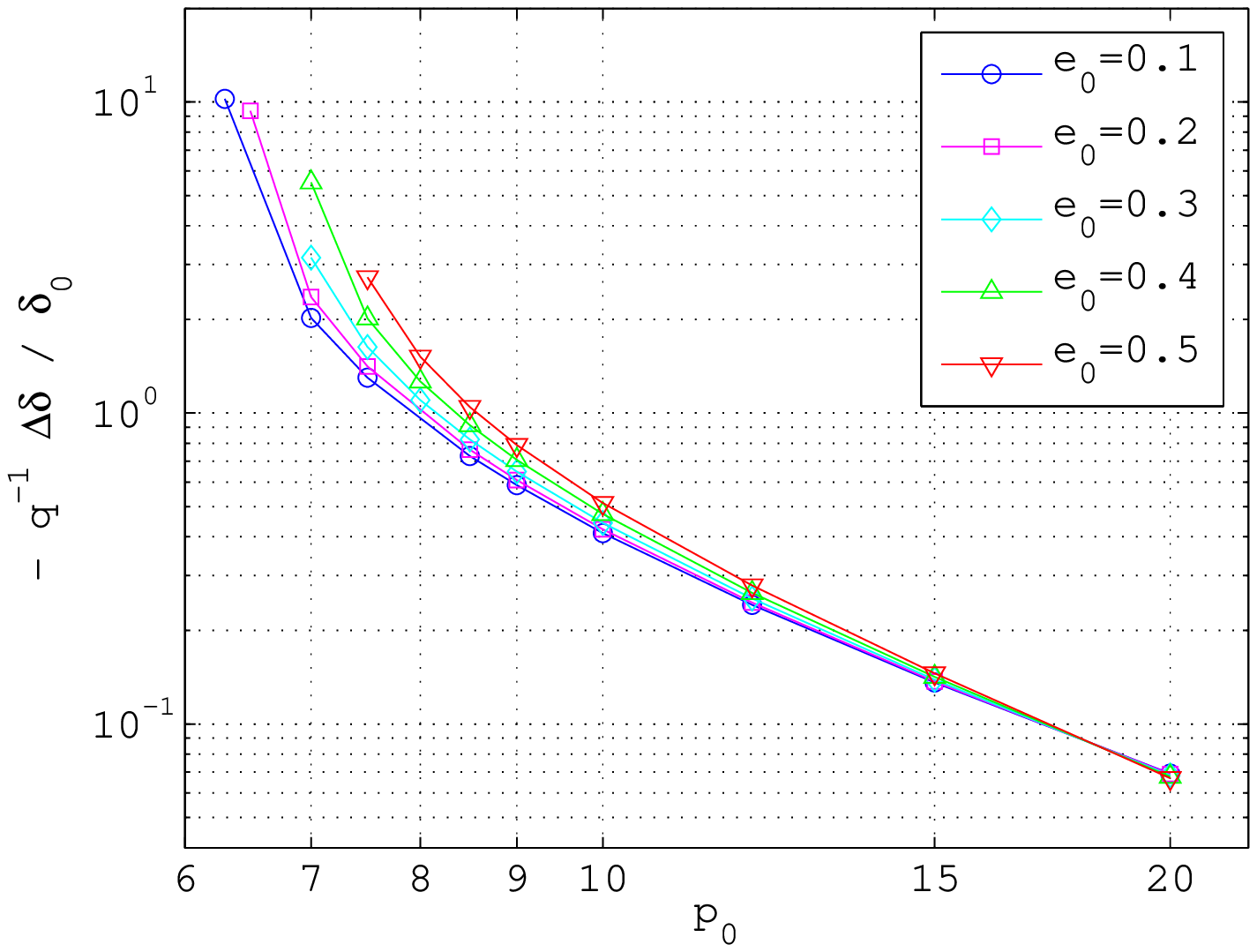}
\caption{Plot of some of the numerical data presented in Table \ref{table:delta_data1}. The left and right panels show, respectively, the absolute and relative GSF corrections, $\Delta\delta$ and $\Delta\delta/\delta_0$ (divided by the mass ratio $q\equiv \mu/M$), as functions of $p_0$ for a variety of eccentricities $e_0$. The data are shown on a log-log scale, and we have in fact plotted $-\Delta\delta$ and $-\Delta\delta/\delta_0$ since $\Delta\delta$ itself is negative (while $\delta_0$ is positive). Note $\Delta\delta$ depends very weakly on $e_0$ at large $p_0$, where it appears to fall off with a power-law $\propto p_0^{-4}$. The background advance $\delta_0$ falls off as $\sim 3p_0^{-1}$ [recall Eq.\ (\ref{delta_largep})]. The dashed line in the left panel is a reference line $\Delta\delta=-2000q/p_0^4$.
}
\label{fig:delta}
\end{figure}

Tables \ref{table:delta_data1}, \ref{table:delta_data2} and Figure \ref{fig:delta} display a sample of numerical results for $\Delta\delta$ over the parameter space of $p$ and $e$. The tables also compare $\Delta\delta$ to the `background' (geodesic) advance $\delta_0$, as derived from Eq.\ (\ref{delta0}) [with Eq.\ (\ref{Phi0})]. We can make the following observations. 
\begin{enumerate}
\item The conservative post-geodesic correction $\Delta\delta$ (in the Lorenz gauge) is {\em negative} for any $p_0,e_0$; it causes {\em recession} of the periastron and  acts to {\em reduce} the rate of periastron precession.
\item At large $p_0$, $\Delta\delta$ exhibits a very weak dependence on $e_0$. This is reminiscent of the behavior of the background advance $\delta_0$; recall Eq.\ (\ref{delta_largep}).
\item At large $p_0$, the magnitude of the GSF correction $\Delta\delta$ seems to fall off faster with $p_0$ than that of the background advance $\delta_0$; the latter [recall Eq.\ (\ref{delta_largep})] falls off as $\propto 1/p_0$ (at fixed $e_0$). Indeed, Fig.\ \ref{fig:delta} suggests the large-$p_0$ behavior
\begin{equation}\label{large_p}
\Delta\delta(p_0\gg 1) \propto p_0^{-4},
\end{equation}
where the proportionality constant ($\sim -2000\mu/M$) is independent of $e_0$.

\item Near the separatrix, the quantity $\Delta\delta$ diverges as $\propto(p_0-6-2e_0)^{-1}$. 
\end{enumerate}

\begin{table}[htb]
\begin{tabular}{llcc}
\hline\hline
$p_0$ & $e_0$ & $(\epsilon/q)\Delta\delta$ & $(\epsilon/q)\Delta\delta/\delta_0$ \\
\hline\hline
$6.301$ & $0.15$ & $-3.58(1)$ & $-0.424(1)$ \\
$6.401$ & $0.20$ & $-3.204(6)$ & $-0.4265(8)$ \\
$6.451$ & $0.225$ & $-3.073(5)$ & $-0.4293(7)$ \\
$6.501$ & $0.250$ & $-2.967(4)$ & $-0.4328(7)$ \\
$6.551$ & $0.275$ & $-2.879(5)$ & $-0.4368(7)$ \\
$6.601$ & $0.30$ & $-2.808(4)$ & $-0.4414(7)$ \\
$6.651$ & $0.325$ & $-2.750(3)$ & $-0.4466(5)$ \\
$6.751$ & $0.375$ & $-2.661(3)$ & $-0.4580(5)$ \\
$6.801$ & $0.40$ & $-2.629(2)$ & $-0.4645(4)$ \\
$6.851$ & $0.425$ & $-2.603(2)$ & $-0.4712(4)$ \\
$6.901$ & $0.45$ & $-2.582(2)$ & $-0.4782(5)$ \\
$6.951$ & $0.475$ & $-2.567(2)$ & $-0.4858(4)$ \\
$7.001$ & $0.50$ & $-2.556(2)$ & $-0.4937(3)$ \\
\hline
$6.110$ & $0.05$ & $-5.98(3)$ & $-0.659(4)$ \\
$6.160$ & $0.075$ & $-4.91(2)$ & $-0.625(3)$ \\
$6.210$ & $0.10$ & $-4.29(1)$ & $-0.607(2)$ \\
$6.260$ & $0.125$ & $-3.89(1)$ & $-0.598(2)$ \\
\hline\hline
\end{tabular}
\caption{Additional numerical data for $\Delta\delta$, exploring the near-separatrix regime. The structure of this table is similar to that of Table \ref{table:delta_data1}, but for convenience we have multiplied here the values of $\Delta\delta$ and $\Delta\delta/\delta_0$ by the separatrix distance $\epsilon\equiv p_0-6-2e_0$. The entries in the upper part of the table correspond to $\epsilon=0.001$, and those in the lower part to  $\epsilon=0.01$.
}
\label{table:delta_data2}
\end{table}

The last of the above observations is particularly interesting. Recall that the background advance $\delta_0$ diverges at the separatrix only logarithmically with $p_0-6-2e_0$ [for fixed $e_0>0$; see Eq.\ (\ref{delta_separatrix})]. This means that, for any given mass ratio $q$, small as we wish, the GSF correction $\Delta\delta$ becomes comparable in magnitude to the background advance $\delta_0$ at some separatrix distance $p_0-6-2e_0\sim O(q)$, and would in fact become a dominant effect (giving rise to a net {\em recession} of the periastron!) at distances smaller still. Inspecting the data in Table \ref{table:delta_data1} we find, for example, that with $q=1/2075$ the conservative GSF effect becomes comparable in magnitude (and opposite in sign) to the background effect for a near-separatrix geodesic with $(p_0,e_0)=(7,0.4999)$. In such a situation one would no longer trust our perturbative treatment, and the leading-order GSF would seem to be of a limited utility.

It should be explained immediately, however, that we do not expect the above situation to realize itself in actual, inspiralling systems. In the physical system, radiation reaction ``smears'' the transition regime across an interval $\Delta r=O(q^{2/5})$ \cite{Buonanno:2000ef,Ori:2000zn}, over which the orbit evolves quickly and is highly nongeodesic. For relevant (small) mass ratios $q$, the above condition $p_0-6-2e_0\sim O(q)\ll \Delta r$ sets the orbit to lie well within this transition regime, where the quantity $\Delta\delta$, defined along closed geodesics, is no longer physically meaningful.  We may ask, conversely, what the magnitude of $\Delta\delta/\delta_0$ is at the onset of the transition regime, i.e., for $p_0-6-2e_0\sim q^{2/5}$. The data in Table \ref{table:delta_data1} indicate that this magnitude may range from about $10^{-2}$ for $q=10^{-3}$ to about $10^{-4}$ for $q=10^{-6}$, and is therefore always quite small in the relevant range of mass ratios.  

\subsection{Effect of dissipation}\label{subsec:diss}

So far we have been making the (unphysical) assumption that the particle experiences no dissipative forces and that, consequently, the orbit remains strictly periodic. In reality, of course, the dynamics is nonconservative and the orbital parameters (e.g., $E$ and $L$) drift secularly in time under the effect of the dissipative piece of the GSF. In a regime where the drift is sufficiently slow (``adiabatic'') we may still define the periastron advance by referring to the $\varphi$-phase accumulated over the period of time between two successive periastron passages (minima of $r$), notwithstanding the fact that two such periastra would occur at slightly different radii [if the radiation-reaction timescale is $E/\dot{E}\sim O(M^2/\mu)$, which we assume here, the two successive periastron radii would differ by a small amount $\propto \mu$]. Let us denote by $\delta_{\rm true}$ the ``true'' periastron advance, defined between two periastron passages of the actual evolving orbit, taking into account all $O(\mu)$ effects of the GSF, both conservative and dissipative (below we give a more precise definition of  $\delta_{\rm true}$). Here (and in Appendix \ref{AppB}) we will argue that, through $O(\mu)$, {\em the quantity  $\delta_{\rm true}$ is, in fact, given by the ``conservative-only'' advance $\delta$, calculated along a suitable ``average'' geodesic orbit}. 

Let us make this statement more precise. First, we need a notion of ``slowly evolving'' $p$ and $e$. For our purpose, it would suffice to concentrate on a particular radial cycle of the evolving orbit, between two apastron passages at times (say) $t_1$ and $t_2$ ($t_2>t_1$). We denote the radius of the (full GSF-perturbed) evolving orbit by $\tilde r(t)$, so that $\tilde r_1\equiv \tilde r(t_1)$ and $\tilde r_2\equiv \tilde r(t_2)$ are two consecutive apastron radii of the true orbit. We assume that at times $t_{1}$ and $t_{2}$ the true orbit is tangent to periodic orbits with parameters $(p_{1},e_{1})$ and $(p_{2},e_{2})$, respectively, which are geodesics perturbed by the conservative piece of the GSF only (in Lorenz gauge, for concreteness). Through $O(\mu)$, then, the ``slowly evolving'' $p$ and $e$ are given, for $t_1\leq t\leq t_2$, by the linear interpolation 
\begin{equation}\label{hatp}
\tilde p(t)=p_1+\frac{p_2-p_1}{t_2-t_1}(t-t_1) +O(\mu^2),
\end{equation}
and similarly for $\tilde e(t)$. Note that terms quadratic (and higher) in $t-t_1$ are also higher order in $\mu$, and so are any conservative GSF corrections to the term linear in $t-t_1$---both types of higher-order corrections can be neglected in our discussion.

We further assume that the solution to the equations of motion for the true, evolving orbit can be obtained [through $O(\mu)$] from the corresponding ``conservative'' solution (i.e., assuming only the conservative piece of the GSF is at play) via the simple replacement $(p,e)\to[\tilde p(t),\tilde e(t)]$. Hence, in particular, we assume [recalling Eq.\ (\ref{tr})]
\begin{equation}\label{hatrofchi}
\tilde r(\tilde\chi)=\frac{\tilde p(t(\tilde\chi))M}{1+\tilde e(t(\tilde\chi))\cos\tilde\chi},
\end{equation}
where $\tilde\chi$ is a certain parameter along the evolving orbit, and the relation $t(\tilde\chi)$ is obtained (in principle) by replacing $(p,e)\to[\tilde p(t),\tilde e(t)]$ and $\chi\to\tilde\chi$ in the ``conservative'' expression for $dt/d\chi$ and integrating with the initial condition $t(\tilde\chi_1)=t_1$. Here $\tilde\chi_1$ is the value of $\tilde\chi$ at the first apastron, determined from the conditions $\tilde\chi_1\to -\pi$ for $\mu\to 0$ together with $\left(d\tilde r/d\tilde\chi\right)_{\tilde\chi=\tilde\chi_1}=0$ through $O(\mu)$ (we determine the value $\tilde\chi_1$ explicitly in Appendix \ref{AppB}). 
Note the form $\tilde r=\tilde r[\tilde\chi;\tilde p(t),\tilde e(t)]$, which reflects the two-timescale dependence of the orbital radius: $\tilde r$ depends on a ``fast'' variable $\tilde\chi$ (or $t$), as well as on the slowly varying parameters $\tilde p$ and $\tilde e$. Writing $\tilde r(\tilde\chi)$ as in Eq.\ (\ref{hatrofchi}) amounts to assuming that, at any fixed time $t=t_0$ ($t_1\leq t_0\leq t_2$), the relation $\tilde r=\tilde r[\tilde\chi;\tilde p(t_0),\tilde e(t_0)]$ describes a ``conservative'' orbit of constant parameters $ p=\tilde p(t_0)$ and $e=\tilde e(t_0)$, tangent to the true, evolving orbit.

In a similar fashion, we write for the $\varphi$-phase along the evolving orbit
\begin{equation}\label{hatdphidchi}
\frac{d\tilde\varphi}{d\tilde\chi} \equiv \tilde\varphi_{\chi}(\tilde\chi)=\tilde\varphi_{\chi}[\tilde\chi;\tilde p(t(\tilde\chi)),\tilde e(t(\chi))],
\end{equation}
where the right-hand side is obtained from Eq.\ (\ref{tildeomega}) by replacing $(p,e)\to[\tilde p(t),\tilde e(t)]$. The total $\varphi$-phase accumulated over one radial period of the evolving orbit ($=t_2-t_1$) is then given by  
\begin{equation}\label{Omegaf_true}
\Phi_{\rm true}=\int_{\tilde\chi_1}^{\tilde\chi_2}\tilde\varphi_{\chi}[\tilde\chi;\tilde p(t(\tilde\chi)),\tilde e(t(\tilde\chi))] d\tilde\chi,
\end{equation}
where, recall, $\tilde\chi_1=\tilde\chi(t_1)$, and we also denoted $\tilde\chi_2\equiv\tilde\chi(t_2)$. The ``true'' periastron advance (per radian) of the evolving orbit between $t_1$ and $t_2$ is given by 
\begin{equation}\label{delta_true}
\delta_{\rm true}=(2\pi)^{-1}\Phi_{\rm true}  - 1.
\end{equation}

With these definitions, our claim is that the true advance $\delta_{\rm true}$, which accounts at $O(\mu)$ for both conservative and dissipative effects of the GSF, can be computed through 
\begin{equation}\label{true}
\delta_{\rm true} = \delta(\bar p,\bar e) +O(\mu^2),
\end{equation}
where 
\begin{equation}\label{av}
\bar p\equiv \frac{1}{2}(p_1+p_2), \quad\quad \bar e\equiv \frac{1}{2}(e_1+e_2).
\end{equation} 
The quantity $\delta(\bar p,\bar e)$ on the right-hand side of Eq.\ (\ref{true}) is the ``conservative-only'' advance, calculated along the conservative GSF-perturbed orbit with ``average'' parameters $(\bar p,\bar e)$. 
It is not difficult to convince oneself of the validity of Eq.\ (\ref{true}); a proof of this relation is presented in Appendix \ref{AppB}. Note that the ``average'' geodesic for use in Eq.\ (\ref{true}) can alternatively be defined through the average values of $E$ and $L$, or of any other pair of slowly varying orbital parameters [because the secular drift in any such parameters over a radial period would be linear in $t$ through $O(\mu)$]. Note also that, through $O(\mu)$, $\bar p,\bar e$ are the momentary parameter values at the {\em periastron} of the evolving orbit. 

We point out that the relation (\ref{true}) is gauge invariant in an obvious sense, since we are considering here the full [$O(\mu^0)+O(\mu)$] periastron advance without breaking it up into ``background'' and ``GSF-correction'' pieces as we have done before. One may, of course, also define the (gauge-dependent) GSF correction in $\delta_{\rm true}$ via $\Delta\delta_{\rm true}\equiv \delta_{\rm true}- \delta_0(\bar p,\bar e)$, which, by virtue of Eqs.\ (\ref{def:deltaSF}) and (\ref{true}), would relate to the conservative-only correction $\Delta\delta$ via
\begin{equation}\label{Deltatrue}
\Delta\delta_{\rm true} = \Delta\delta(\bar p,\bar e) +O(\mu^2).
\end{equation}

Equations (\ref{true}) and (\ref{Deltatrue}) suggest a way in which our results for the conservative-only periastron advance $\delta$ may be relevant to actual, evolving orbits: One only needs to {\em reinterpret} the quantity $\delta(p,e)$ as the advance (per radian) of an adiabatically evolving orbit, defined [via Eqs.\ (\ref{Omegaf_true}), (\ref{delta_true})] between two periastra whose average parameter values are $(p,e)$. This interpretation [which, we recall, is only valid through $O(\mu)$] may form a basis for comparison between our GSF precession data and results from fully nonlinear numerical-relativistic simulations of binary inspirals, once the latter are available for sufficiently small mass ratios. This interpretation also suggests a way in which our results for the conservative correction $\Delta\delta$ might be incorporated into one of the existing approximate (PN/perturbative) frameworks for computing the orbital evolution in extreme-mass-ratio systems \cite{Gair:2005ih,Sundararajan:2008zm,Gair:2010iv}, which so far do not account for the conservative effects of the GSF. (Any such application of our results will, however, need to deal cautiously with the gauge dependence of $\Delta\delta$.)

\section{Gauge-invariant effect of the conservative GSF}\label{Sec:U}

The quantity $\Delta\delta(p,e)$ considered above is gauge dependent, and as such it is of a limited utility. Our ambition in this section is to devise a {\em gauge-invariant} measure of the $O(\mu)$ conservative GSF effect on eccentric geodesics. As already mentioned, while $\delta$ itself is gauge invariant (as it is constructed from the two fundamental frequencies of the orbit, which are gauge invariant), the $O(\mu)$ functional relation $\Delta\delta(p,e)$ is not invariant since the parameters $p,e$ are gauge dependent. A gauge-invariant parametrization of the eccentric orbits is provided by the pair $(\Omegaf,\Omegar)$, but the relation $\delta(\Omegaf,\Omegar)$ is trivial [recall Eq.\ (\ref{tdelta})] and gives no information about the GSF. What is required is a {\em third}, independent gauge-invariant quantity (call it $G$) which depends in a nontrivial way on the two invariant frequencies. The relation $\Delta G(\Omegaf,\Omegar)\equiv G(\Omegaf,\Omegar)-G_0(\Omegaf,\Omegar)$ [where, recall, $ G(\cdot)$ and $G_0(\cdot)$ denote the functional relations with and without the GSF, respectively] would then provide a gauge-invariant measure of the GSF effect. 

The quest for such a function $G$ is motivated by the wish to identify a common reference point for comparison between calculations of the GSF held in different gauges \cite{Sago:2008id}, and also to facilitate comparison with results from PN theory. We therefore remain mindful that our $G$ would need to be readily accessible to both GSF and PN treatments. In what follows we propose such a gauge-invariant $G$ and use our code to compute the GSF correction $\Delta G(\Omegaf,\Omegar)$, ready for comparison with results from other methods when these become available.

\subsection{``Physically acceptable'' gauge transformations}\label{subsec:gauge}

First, we need to make precise the meaning of gauge invariance in our problem. An $O(\mu)$ gauge transformation 
\begin{equation}\label{xi}
x^{\alpha}\to x^{\alpha}-\xi^{\alpha}
\end{equation}
changes the physical (retarded) metric perturbation $h_{\alpha\beta}$ associated with particle by an amount 
\begin{equation}\label{h_trans}
\delta_{\xi} h_{\alpha\beta}=\xi_{\alpha;\beta}+\xi_{\beta;\alpha},
\end{equation}
where a semicolon denotes covariant differentiation with respect to the background (Schwarzschild) metric. In the Detweiler--Whiting interpretation \cite{Detweiler:2002mi}, the GSF is exerted by a certain smooth perturbation $h^R_{\alpha\beta}$ derived from $h_{\alpha\beta}$. This so-called ``R-field'' (which is defined only in the local neighbourhood of the particle) transforms under $\xi^{\alpha}$ in the same way as $h_{\alpha\beta}$ \cite{Barack:2001ph,Barack:2009ux}:
\begin{equation}\label{hR_trans}
\delta_{\xi} h^R_{\alpha\beta}=\xi_{\alpha;\beta}+\xi_{\beta;\alpha}.
\end{equation}
It is usually desirable to work in a gauge in which the metric perturbation ($h_{\alpha\beta}$ or $h^R_{\alpha\beta}$) correctly reflects the underlying symmetry of the physical system. In the case of eccentric geodesics, an essential requirement is for the metric perturbation to respect the periodicity of the orbit, i.e., to exhibit a bi-periodic spectrum with fundamental frequencies $\Omega_r$ and $\Omega_\varphi$. In particular, when evaluated along the geodesic orbit (say, as a function of proper-time $\tau$), the field $h^R_{\alpha\beta}$ need be periodic with a $\tau$-period $\calT$ corresponding to the radial $t$-period $T$ (this forbids, for example, a spurious secular growth of the perturbation over time). We write this periodicity condition as
\begin{equation}\label{hR_period}
h^R_{\alpha\beta}[x_p^{\mu}(\tau)]=h^R_{\alpha\beta}[x_p^{\mu}(\tau+\calT)]
\end{equation}
(for any proper-time $\tau$ along the orbit), where $x^{\mu}=x^{\mu}_{p}(\tau)$ describes the geodesic orbit. In the special case of a circular orbit, the condition (\ref{hR_period}) is replaced with the requirement of helical symmetry \cite{Sago:2008id}. We require the R-field in any ``physically acceptable'' gauge to satisfy (\ref{hR_period}). 

In discussing gauge invariance we wish to restrict attention to the class of transformations $\xi^{\alpha}$ that take one ``physically acceptable'' perturbation to another. Such transformations need to preserve, in particular, the periodicity of the metric perturbation. This can be achieved by requiring that the generators $\xi^{\alpha}$ themselves are bi-periodic functions of $t$ (with frequencies $\Omega_r$ and $\Omega_\varphi$), so that, in particular,
\begin{equation}\label{xi_period}
\xi^{\alpha}[x^{\mu}_p(\tau)]=\xi^{\alpha}[x^{\mu}_p(\tau+{\cal T})]
\end{equation}
for any $\tau$. In fact, to this class of transformations we may add $t$-translations of the form $\xi^{\alpha}\propto t \delta^{\alpha}_t$, which are themselves nonperiodic but do not interfere with the periodicity of the metric perturbation; we shall come back to this type of gauge displacements later in our discussion. For now, however, let us restrict attention to the class of gauge transformations satisfying the periodicity condition (\ref{xi_period}). We shall refer to any quantity defined along the orbit as ``gauge invariant'' if it is invariant under any gauge transformation within this class. \footnote{
The technical notion of gauge invariance in the case of a quantity that is only defined along the orbit is subtly different from that of a field like $h^{R}_{\alpha\beta}$. The operator $\delta_\xi$ in Eq.\ (\ref{hR_trans}), for instance, measures the difference in the {\it functional form} of $h^{R}_{\alpha\beta}(x)$ as a result of the transformation (\ref{xi}), i.e., the difference between the `new' and `old' fields evaluated at the same coordinate value (and hence, generally, at different physical points). On the other hand, for quantities like $u^{\alpha}$ or $\Omega_{\varphi}$, which are only defined along the orbit, one considers the difference between the `new' and `old' quantities evaluated {\em at the same physical point} (same value of $\tau$), which in general takes different coordinate values before and after the transformation. Ref.\ \cite{Shah:2010bi} discusses this point in detail.}  

Let us consider, for example, the frequencies $\Omega_r$ and $\Omega_{\varphi}$, already described above as ``gauge invariant''. It is readily seen that the $t$-period $T$ is formally invariant under periodic gauge transformations: Changing integration variables in Eq.\ (\ref{Omegar0}) gives
\begin{equation}\label{T_inv1}
T=\int_{0}^{\calT}\frac{dt}{d\tau}d\tau 
\end{equation}
(taking $\tau=0$ at periastron without loss of generality), which, under the transformation (\ref{xi}), is modified by an amount 
\begin{equation}\label{T_inv2}
\delta_\xi T=-\int_{0}^{\calT}\frac{d\xi^t}{d\tau}d\tau=-\xi^t(\calT)+\xi^t(0)=0, 
\end{equation}
following from the periodicity condition (\ref{xi_period}) [we hereafter use $\xi^{\alpha}(\tau)$ as a shorthand for $\xi^{\alpha}(x^{\mu}_p(\tau))$]. Hence, the radial frequency $\Omega_r=2\pi/T$ too is gauge invariant. Similarly, we have for the azimuthal frequency 
\begin{equation}\label{Omegaphi_inv1}
\Omega_{\varphi}=\frac{1}{T}\int_{0}^{\calT}\frac{d\varphi}{d\tau}d\tau ,
\end{equation}
giving
\begin{equation}\label{Omegaphi_inv2}
\delta_\xi \Omega_{\varphi}=-\frac{1}{T}\int_{0}^{\calT}\frac{d\xi^\varphi}{d\tau}d\tau=-T^{-1}[\xi^\varphi(\calT)-\xi^\varphi(0)]=0, 
\end{equation}
by virtue of the gauge invariance of $T$ and the periodicity condition (\ref{xi_period}). This confirms that $\Omega_\varphi$ too is gauge invariant. Note that the formal invariance of $\Omega_r$ and $\Omega_\varphi$ is a direct consequence of the periodicity condition (\ref{xi_period}); in general, the frequencies will not remain invariant under gauge transformations that fail to satisfy (\ref{xi_period}). 

\subsection{Generalized redshift invariant}

Detweiler \cite{Detweiler:2005kq} first pointed out that, in the case of a circular orbit, the quantity $U\equiv u^t$ (i.e., the conservative GSF-perturbed $t$ component of the particle's four-velocity) is invariant under gauge transformations that respect the helical symmetry of the black hole--particle configuration. A possible physical interpretation of $U$ as an observable measure of gravitational redshift is discussed in Ref.\ \cite{Detweiler:2008ft}. The gauge-invariant relation $U(\Omega_\varphi)$ was later utilized for comparing between GSF results in different gauges \cite{Sago:2008id,Shah:2010bi} and between GSF and PN results \cite{Detweiler:2008ft,Blanchet:2009sd,Blanchet:2010zd}.

Here we propose a natural generalization of the redshift invariant to the case of eccentric orbits. Our proposed invariant is simply the $\tau$-average of $U$ over a radial period, which is also, more simply, the ratio between the $t$ and $\tau$ periods:
\begin{equation}\label{def:U}
\U_{\tau} \equiv
\frac{1}{\calT}\int_0^{\calT}\frac{d{t}}{d{\tau}}d\tau=\frac{T}{\calT}.
\end{equation}
The quantity $\U_{\tau}$ is obviously gauge invariant, since both $T$ and $\cal T$ are invariant. (We could as well choose $\cal T$ as our third invariant; we prefer $\U_{\tau}$ because it is dimensionless and because is reduces to the standard redshift invariant in the circular-orbit limit.) 

As in \cite{Sago:2008id}, we make two small adjustments to our invariant, to better fit it for comparison with PN calculations. First, we replace the proper-time $\tau$ (which is defined with respect to the background metric $g_{\alpha\beta}$) with the proper-time $\ttau$ defined with respect to the perturbed metric $g_{\alpha\beta}+h_{\alpha\beta}^R$. The two proper times are related, through $O(\mu)$, via
\begin{equation}
\frac{d\tau}{d\ttau} = 1+H^R,
\end{equation}
where 
\begin{equation}
H^R\equiv \frac{1}{2} h_{\alpha\beta}^R u^\alpha u^\beta,
\end{equation}
with the R-field perturbation $h_{\alpha\beta}^R$ evaluated at the particle.
(Here the four-velocity $u^\alpha$ may be defined with respect to either $\tau$ or $\tilde\tau$; the difference would affect $H^R$ only at sub-leading order in $\mu$, which we neglect in our treatment.)
The R-field combination $H^R$ can be constructed directly from the full (retarded) metric perturbation using a certain regularization procedure, which we describe in Appendix \ref{App:H^R}. The appendix also details the numerical computation of $H^R$ in practice, using our code.   

The second adjustment is a normalization of the time coordinate $t$, motivated as follows. As first mentioned in \cite{Barack:2005nr} and discussed in length in \cite{Sago:2008id,Damour:2009sm,Barack:2010ny}, the physical metric perturbation component $h_{tt}$, in the Lorenz-gauge, has the peculiarity that it does not vanish at $r\to\infty$ but rather approaches a constant nonzero value (which depends only on the orbital parameters). This behavior is entirely attributed to the static piece of the mass monopole perturbation, and thus the asymptotic value of $h_{tt}$ does not depend on the angular direction even for eccentric orbits. To remove this gauge artifact and facilitate comparison with PN theory (where a more suitable ``asymptotically flat'' time coordinate is used), we introduce the normalized time coordinate 
\begin{equation}\label{hatt}
\hat t=(1+\alpha)t,
\end{equation}\label{alpha}
where $\alpha=\alpha(p,e)$ is given by
\begin{equation}\label{alpha}
 \alpha = -\frac{1}{2}h_{tt}(r\rightarrow \infty).
\end{equation}
Then, through $O(\mu)$, we have $g_{\hat t \hat t}+h_{\hat t \hat t}\to -1$ as $r\to\infty$, as desired. In the circular-orbit case one finds \cite{Barack:2005nr} $\alpha(e=0) = \mu[r_\circ(r_\circ-3M)]^{-1/2}$, where $r=r_\circ$ is the radius of the orbit. For eccentric orbits we do not have an analytic expression for $\alpha$, but its numerical value can be extracted from our numerical solutions for the Lorenz-gauge perturbation, using Eq.\ (\ref{alpha}). Note that the transformation $t\to \hat t$ amounts to an $O(\mu)$ gauge transformation (\ref{xi}) with $\xi^{\alpha}= -\alpha t \delta^{\alpha}_t \equiv \hat\xi^{\alpha}$. This transformation from the Lorenz-gauge time $t$ to the ``asymptotically flat'' time $\hat t$ does not spoil the periodicity of the metric perturbation, but its generator $\hat\xi^{\alpha}$ does not respect the periodicity condition (\ref{xi_period}). As a result, the transformation does not leave $\U_\tau$ invariant; in fact, one finds $\delta_{\hat\xi}\U_\tau=\alpha\U_\tau$.

With the above two adjustments, we redefine our generalized redshift invariant as
\begin{equation}\label{def:hatU}
\hatU \equiv
\frac{1}{\tilde\calT}\int_0^{\tilde\calT}\frac{d{\hat t}}{d{\ttau}}d\ttau=\frac{\hat T}{\tilde\calT},
\end{equation}
where 
\begin{equation}\label{hatT}
\hat T=(1+\alpha)T, \quad\quad
\tilde\calT=\int_0^{T}(1-H^R)\left(\frac{dt}{d\tau}\right)^{-1}dt
\end{equation}
are [through $O(\mu)$] the radial periods measured in time $\hat t$ and proper time $\ttau$, respectively.

\subsection{The geodesic limit of $\hatU$}

At the limit $\mu\to 0$ the quantities $\hatU$ and $\U_{\tau}$ coincide, and are given by
\begin{equation}\label{def:U0}
\hatUbg=\U_{\tau 0} = \frac{T_0}{\calT_0},
\end{equation}
where $T_0=T_0(p_0,e_0)$ and $\calT_0=\calT_0(p_0,e_0)$ are the geodesic radial periods measured in $t$ and $\tau$, respectively. The period $T_0$ is computed via Eqs.\ (\ref{Omegar0}), which, for easy reference, we reproduce here in explicit form:
\begin{equation}\label{T0exp}
T_0(p,e)= 2Mp^2[(p-2)^2-4e^2]^{1/2}\int_{0}^{\pi}\frac{(p-6-2e\cos\chi)^{-1/2}}{(p-2-2e\cos\chi)(1+e\cos\chi)^2}\, d\chi.
\end{equation} 
The proper-time period $\calT_0$ is obtained by integrating $d\tau/d\chi$ given in Eq.\ (\ref{dtaudchi}):
\begin{equation}\label{Tau0exp}
\calT_0(p,e)=2Mp^{3/2}(p-3-e^2)^{1/2}\int_0^{\pi}\frac{(p-6-2e\cos\chi)^{-1/2}}{(1+e\cos\chi)^2 } d\chi.
\end{equation}

\subsection{Gauge-invariant parametrization of the orbit}

As already discussed, we are aiming to replace the $p,e$ parametrization of the orbit with a gauge-invariant one, based on the (perturbed) fundamental frequencies $\Omegar,\Omegaf$. More precisely, we wish to work with the ``normalized'' frequencies, defined with respect to time $\hat t$. These are simply related to the original $t$-frequencies via
\begin{equation}\label{homega}
\hOmegar=(1-\alpha)\Omegar, \quad\quad
\hOmegaf=(1-\alpha)\Omegaf,
\end{equation} 
valid through $O(\mu)$. Rather than using the frequencies $\hOmegar,\hOmegaf$ themselves, we find it convenient to introduce a new pair of gauge-invariant parameters, denoted $\pI$ and $\eI$, which are obtained by inverting
\begin{equation}\label{invert}
\hOmegar=\frac{2\pi}{T_0(\pI,\eI)},\quad\quad
\hOmegaf=\frac{\Phi_0(\mathsf{\pI,\eI})}{T_0(\pI,\eI)},
\end{equation}
where $T_0$ and $\Phi_0$ are the {\em geodetic} relations, given respectively in Eqs.\ (\ref{T0exp}) and (\ref{Phi0}) (with the replacement $p,e\to\pI,\eI$), and, recall, $\hOmegar$ and $\hOmegaf$ are the {\em GSF-perturbed} frequencies. The quantities $\pI$ and $\eI$ are natural gauge-invariant notions of ``semilatus rectum'' and ``eccentricity'', in much the same way that the standard quantity $x^{-1}=(M\Omegaf)^{-2/3}$ (see, e.g., \cite{Barack:2010ny}) is a natural gauge-invariant notion of radius in the circular-orbit case. 

It is important to recall, however, that the relations $\hOmegar(\pI,\eI)$ and $\hOmegaf(\pI,\eI)$ in Eq.\ (\ref{invert}) are {\em not} bijective, and thus cannot be inverted without a suitable restriction of the domain---we remind the reader of the discussion at the end of Sec.\ \ref{subsec:prelim} and in Appendix \ref{App:sing}. The inverse relations $\pI(\hOmegar,\hOmegaf)$ and $\eI(\hOmegar,\hOmegaf)$ are nonetheless well defined in each of the domains $\pI>\pI_s(\eI)$ and $\pI<\pI_s(\eI)$ (cf.\ Fig.\ \ref{fig:sing}) separately. In what follows we will assume that the domain [$\pI>\pI_s(\eI)$ or $\pI<\pI_s(\eI)$] has been prespecified and that $\pI(\hOmegar,\hOmegaf)$ and $\eI(\hOmegar,\hOmegaf)$ are the uniquely determined values corresponding to that domain.

Even in a suitably restricted domain, it is not possible to invert Eq.\ (\ref{invert}) in explicit form to obtain $\pI(\hOmegar,\hOmegaf)$ and $\eI(\hOmegar,\hOmegaf)$. However, given the GSF in a particular gauge, we may express the invariants $\pI,\eI$ through $O(\mu)$ in terms of the gauge-dependent parameters $e,p$, using the linear variation formulas
\begin{eqnarray} \label{pI}
\pI &=&  
p
+ \frac{\partial \pI}{\partial \hOmegar}\bigg|_0
  \Delta\hOmegar
+ \frac{\partial \pI}{\partial \hOmegaf}\bigg|_0
  \Delta\hOmegaf
+ O(\mu^2), \\
\eI &=&  \label{eI}
e
+ \frac{\partial \eI}{\partial \hOmegar}\bigg|_0
  \Delta\hOmegar
+ \frac{\partial \eI}{\partial \hOmegaf}\bigg|_0
  \Delta\hOmegaf
+ O(\mu^2).
\end{eqnarray}
Here the partial derivatives can be evaluated by inverting the transformation matrix $\partial(\hOmegar,\hOmegaf)/\partial(\pI,\eI)$, which itself can be computed numerically (for given $p,e$) based on Eq.\ (\ref{invert}) with (\ref{T0exp}) and (\ref{Phi0}) (replacing $\pI,\eI\to p,e$). The subscript `$0$' indicates that these partial derivatives are to be evaluated at the geodesic limit. The $O(\mu)$ quantities $\Delta\hOmegar$ and $\Delta\hOmegaf$ are the GSF corrections to the corresponding frequencies, defined as
\begin{equation}
\Delta\hOmegar(p,e)\equiv \hOmegar(p,e) - \Omega_{r0}(p,e), \quad\quad
\Delta\hOmegaf(p,e)\equiv \hOmegaf(p,e) - \Omega_{\varphi0}(p,e).
\end{equation}
Unfortunately, the inversion formulas (\ref{pI}) and (\ref{eI}) become meaningless along the singular curve $\pI=\pI_s(\eI)$, where the transformation matrix $\partial(\hOmegar,\hOmegaf)/\partial(\pI,\eI)$ is singular. In the following analysis we will therefore ``keep away'' from parameter-space points that lie directly on the singular curve.

To calculate $\pI(p,e)$ and $\eI(p,e)$ in Eqs.\ (\ref{pI}) and (\ref{eI}) we need explicit expressions for $\Delta\hOmegar$ and $\Delta\hOmegaf$ in terms of $p,e$ and the GSF. These are given, through $O(\mu)$, by
\begin{eqnarray}\label{DeltaOmegar}
\Delta\hOmegar &=&
- \Omega_{r0}\left(\alpha+\frac{\Delta T}{T_0}\right) ,
\\
\Delta\hOmegaf &=& \label{DeltaOmegaf}
-\Omega_{\varphi0}\left(\alpha-\frac{\Delta\Phi}{\Phi_0}+\frac{\Delta T}{T_0}\right),
\end{eqnarray}
where $\Delta T(p,e)\equiv T(p,e)-T_0(p,e)$ and $\Delta\Phi(p,e)\equiv \Phi(p,e)-\Phi_0(p,e)$ are the $O(\mu)$ GSF corrections to $T$ and $\Phi$, themselves given by
\begin{eqnarray}\label{DeltaT}
\Delta T    &=& 2\int_0^{\pi} \Delta t_{\chi}(\chi;p,e) \, d\chi , \\
\Delta \Phi &=& 2\int_0^{\pi} \Delta\varphi_{\chi}(\chi;p,e) \, d\chi = 2\pi \Delta\delta .
\end{eqnarray}
The quantity $\Delta\varphi_{\chi}$, recall, is the GSF correction to $\varphi_{\chi}\equiv d\varphi/d\chi$; it was given explicitly in Eq.\ (\ref{Deltaomega2}) in terms of $e$, $p$ and the GSF quantities ${\cal E}(\chi)$ and ${\cal L}(\chi)$. The quantity $\Delta t_{\chi}$ is the GSF correction to 
\begin{equation}
t_{\chi}\equiv \frac{dt}{d\chi}.
\end{equation}
The evaluation of $\Delta t_{\chi}$ is similar to that of $\Delta\varphi_{\chi}$ (see Sec.\ \ref {subsec:GSFcorrection}). The result is
\begin{eqnarray}\label{Deltagamma}
\Delta t_{\chi} &=&
\frac{M p^{5/2}(p-3-e^2)^{1/2}}{e^2(1+e\cos\chi)^2(p-6-2e\cos\chi)^{3/2}}
\left[
\frac{(p-2-2e)(2+e+e\cos\chi)}{8\cos^2(\chi/2)}{\cal E}(\pi)
- \frac{p-3-e^2+(1+e\cos\chi)^2}{\sin^2\chi}{\cal E}(\chi)
\right]
\nonumber \\ 
&&+
\frac{p(p-3-e^2)^{1/2}[(p-2)^2-4e^2]^{1/2}}
     {e^2(1+e\cos\chi)^2(p-6-2e\cos\chi)^{3/2}}
\left[
\frac{(1+e\cos\chi)^2}{\sin^2\chi}{\cal L}(\chi)
- \frac{(1-e)^2(2+e+e\cos\chi)}{8\cos^2(\chi/2)}{\cal L}(\pi)
\right].
\end{eqnarray}
As with $\Delta\varphi_{\chi}$ [Eq.\ (\ref{Deltaomega2})], here too the evaluation of the expression near the turning point $\chi=0,\pi$ is subtle.  The practical solution proposed in Appendix \ref{AppA} applies here as well.

Let us summarize. Given a parameter-space point $p,e$ and the GSF corresponding to that point, Eqs.\ (\ref{pI}) and (\ref{eI}) are used to construct the two quantities $\pI(p,e)$ and $\eI(p,e)$. The pair $(\pI,\eI)$ constitutes a gauge-invariant parametrization of the eccentric orbits [on each side of the singular curve $\pI=\pI_s(\eI)$].

\subsection{Conservative GSF correction to $\hatU$}

We are now in position to write down a gauge-invariant expression for the post-geodesic correction to $\hatU$:
\begin{equation}\label{DeltaU}
\Delta\hatU \equiv \hatU(\pI,\eI)-\U_{\tau 0}(\pI,\eI).
\end{equation}
Here $\U_{\tau 0}(\pI,\eI)$ represents the geodesic functional relation given in Eq.\ (\ref{def:U0}); it is to be calculated using Eqs.\ (\ref{T0exp}) and (\ref{Tau0exp}) with the arguments $p,e$ replaced with $\pI,\eI$. Since both 
the function $\hatU$ and the parameter pair $\pI,\eI$ are gauge invariant, the $O(\mu)$ quantity $\Delta\hatU$ provides a genuinely invariant description of the GSF effect.

Our next goal is to express Eq.\ (\ref{DeltaU}) in a workable form, i.e., as a function of the parameters $p,e$ and the GSF. To this end, we expand each of the two terms on the right-hand side of Eq.\ (\ref{DeltaU}) about its geodesic value $\U_{\tau 0}(p,e)$ through $O(\mu)$. Starting with $\hatU(\pI,\eI)=\hat T(\pI,\eI)/\tilde\calT(\pI,\eI)$, we obtain
\begin{equation}\label{expand1}
\hatU(\pI,\eI)=\U_{\tau 0}(p,e)\left(1+\frac{\Delta \hat T}{T_0(p,e)}-\frac{\Delta\tilde\calT}{{\calT_0}(p,e)}\right)+O(\mu^2).
\end{equation}
Here $\Delta \hat T$ and $\Delta\tilde\calT$ are the GSF corrections to $\hat T$ and $\tilde\calT$, respectively, which we shall give explicitly below in terms of $p,e$ and the GSF. The second term on the right-hand side of (\ref{DeltaU}) is expanded in the form
\begin{equation}\label{expand2}
\U_{\tau 0}(\pI,\eI)=\U_{\tau 0}(p,e)
+\frac{\partial\U_{\tau 0}}{\partial\pI}\bigg|_0 (\pI-p)
+\frac{\partial\U_{\tau 0}}{\partial\eI}\bigg|_0 (\eI-e)
+O(\mu^2),
\end{equation}
where the subscript `$0$' indicates that the partial derivatives are to be evaluated at $(\pI,\eI)\to(p,e)$.
Substituting Eqs.\ (\ref{expand1}) and (\ref{expand2}) in Eq.\ (\ref{DeltaU}) and using Eqs.\ (\ref{pI}) and (\ref{eI}) we then obtain, neglecting $O(\mu^2)$ terms,
\begin{equation}\label{DeltaUfinal}
\Delta\hatU=\U_{\tau 0}(p,e)\left(\frac{\Delta \hat T}{T_0(p,e)}-\frac{\Delta\tilde\calT}{{\calT_0}(p,e)}\right)
-C_r(p,e) \Delta\hOmegar -C_{\varphi}(p,e)\Delta\hOmegaf
\end{equation}
with 
\begin{eqnarray}\label{C}
C_r(p,e) &=& 
\frac{\partial\U_{\tau 0}}{\partial p}\frac{\partial p}{\partial \Omegar}
+\frac{\partial\U_{\tau 0}}{\partial e}\frac{\partial e}{\partial \Omegar},
\\
C_{\varphi}(p,e) &=& 
\frac{\partial\U_{\tau 0}}{\partial p} \frac{\partial p}{\partial \Omegaf}
+\frac{\partial\U_{\tau 0}}{\partial e} \frac{\partial e}{\partial \Omegaf}.
\end{eqnarray}
The coefficients $C_r$ and $C_{\varphi}$ can be computed (numerically), for any given $p,e$, from the appropriate geodesic expressions: The partial derivatives of $\U_{\tau 0}$ are obtained using Eq.\ (\ref{def:U0}) with (\ref{T0exp}) and (\ref{Tau0exp}), and the partial derivatives $\partial(p,e)/\partial(\Omegar,\Omegaf)$ are computed as explained below Eq.\ (\ref{eI}). Note that in the expressions for $C_r$ and $C_{\varphi}$ we have allowed ourselves to replace $(\hOmegar,\hOmegaf)\to(\Omegar,\Omegaf)$ and $(\pI,\eI)\to(p,e)$, which does not affect $\Delta\hatU$ through $O(\mu^0)$.

Finally, to be able to use Eq.\ (\ref{DeltaUfinal}), we need expressions for the $O(\mu)$ quantities $\Delta \hat T$, 
$\Delta\tilde\calT$, $\Delta\hOmegar$ and $\Delta\hOmegaf$. The latter two have already been given above, in Eqs.\ (\ref{DeltaOmegar}) and (\ref{DeltaOmegaf}). As for $\Delta \hat T$, it follows from Eq.\ (\ref{hatT}) that
\begin{equation}
\Delta \hat T = \alpha T_0(p,e)+\Delta T,
\end{equation}
with $\Delta T$ given in Eq.\ (\ref{DeltaT}) [with (\ref{Deltagamma})]. Lastly, we have, recalling Eq.\ (\ref{hatT}),
\begin{equation}\label{DeltaTau}
\Delta\tilde\calT = 2\int_0^{\pi} \left(\Delta\tau_{\chi}-H^R\tau_{\chi 0}\right) \, d\chi,
\end{equation}
where we introduced the notation 
\begin{equation}
\tau_{\chi}\equiv \frac{d\tau}{d\chi},
\end{equation}
with $\tau_{\chi 0}(p,e)$ being the geodetic limit of $\tau_{\chi}$, and $\Delta\tau_{\chi}(p,e)\equiv \tau_{\chi}(p,e)-\tau_{\chi 0}(p,e)$. The background quantity $\tau_{\chi 0}$ is given explicitly in Eq.\ (\ref{dtaudchi}), and for the perturbed quantity $\tau_{\chi}$ we write 
\begin{equation}
\tau_{\chi}=\frac{dr/d\chi}{dr/d\tau}=\frac{epM |\sin\chi|(1+e\cos\chi)^{-2}}{[E^2(\chi)-V(r(\chi),L(\chi))]^{1/2}},
\end{equation}
proceeding using the method of Sec.\ \ref {subsec:GSFcorrection} to obtain
\begin{eqnarray}\label{Deltatau}
\Delta \tau_{\chi} &=&
\frac{M p^{2}[(p-2)^2-4e^2]^{1/2}(p-3-e^2)}{e^2(1+e\cos\chi)^2(p-6-2e\cos\chi)^{3/2}}
\left[
\frac{{\cal E}(\pi)}{4\cos^2(\chi/2)}-\frac{{\cal E}(\chi)}{\sin^2\chi}
\right]
\nonumber \\ 
&&
-\frac{p^{1/2}(p-3-e^2)}{e^2(p-6-2e\cos\chi)^{3/2}}
\left[
\frac{(1-e)^2(p-2+2e)}{(1+e\cos\chi)^2}\frac{{\cal L}(\pi)}{4\cos^2(\chi/2)}
-(p-2-2e\cos\chi)\frac{{\cal L}(\chi)}{\sin^2\chi}
\right]
\nonumber \\ 
&&
+\frac{Mp^2[(p-2)^2-4e^2]^{1/2}{\cal E}(\pi)-p^{1/2}(1-e)^2(p-2+2e){\cal L}(\pi)}{4e(1+e\cos\chi)^2(p-6-2e\cos\chi)^{1/2}}.
\end{eqnarray}
The method of Appendix \ref{AppA} can again be used to assist in evaluating the last expression near the turning points $\chi=0,\pi$.

Let us summarize the above construction. Our main result is expressed in Eq.\ (\ref{DeltaUfinal}), giving $\Delta\hatU$ in terms of the parameters $p,e$ and the GSF. The various elements of Eq.\ (\ref{DeltaUfinal}) are constructed using Eqs.\ (\ref{def:U0})--(\ref{Tau0exp}), (\ref{DeltaOmegar}), (\ref{DeltaOmegaf}) and (\ref{C})--(\ref{DeltaTau}). Note that since $\Delta\hatU$ is already $O(\mu)$, in Eq.\ (\ref{DeltaUfinal}) we may replace $(p,e)\to (p_0,e_0)$ and calculate the various ingredients of this equation based on GSF data evaluated along geodesic orbits.

The value of $\Delta\hatU$ in Eq.\ (\ref{DeltaUfinal}) {\em does not depend on the gauge} in which the GSF is given, and as such it provides a useful reference for comparison between results from different calculation schemes (PN or perturbative).

\subsection{Numerical results}

Using Eq.\ (\ref{DeltaUfinal}) we have computed $\Delta\hatU$ for a sample of $e,p$ values; the results are displayed in Table \ref{table:U_data}. We have tested these results against weak-field analytic expressions through 1PN, and found a very good agreement---this comparison will be presented in a forthcoming paper \cite{inprep}. In the future, the data could also provide a basis for comparison with other calculations of the GSF (in whatever gauge) when these are available.
\begin{table}[htb]
\begin{tabular}{lllllll}
\hline\hline
$p$ & $e$ & $M\Omega_{r0}\times 100$ & $M\Omega_{\varphi0}\times 100$
    & \quad\ \ $\alpha$ & \quad $\Delta\hatU$ \\
\hline\hline
$6.1$ & $0.02$ & $0.8250131$ & $6.655900$ & $0.2306141$ & $-0.287145(2)$ \\
$6.2$ & $0.05$ & $1.111674$ & $6.527790$ & $0.2263482$ & $-0.279734(2)$ \\
$6.3$ & $0.1$ & $1.264048$ & $6.447631$ & $0.2240162$ & $-0.275394(2)$ \\
$6.4$ & $0.1$ & $1.479287$ & $6.240510$ & $0.2170145$ & $-0.263768(2)$ \\
$6.5$ & $0.1$ & $1.629472$ & $6.068755$ & $0.2112690$ & $-0.254468(2)$ \\
$6.5$ & $0.2$ & $1.447220$ & $6.260868$ & $0.2192002$ & $-0.265667(2)$ \\
$6.7$ & $0.1$ & $1.835687$ & $5.770239$ & $0.2014079$ & $-0.238987(2)$ \\
$6.7$ & $0.2$ & $1.743418$ & $5.800667$ & $0.2037311$ & $-0.240868(2)$ \\
$6.7$ & $0.3$ & $1.532544$ & $5.985119$ & $0.2122705$ & $-0.251457(2)$ \\
$7$ & $0.1$ & $2.019275$ & $5.383470$ & $0.1888463$ & $-0.220085(2)$ \\
$7$ & $0.2$ & $1.953186$ & $5.337941$ & $0.1885223$ & $-0.217914(2)$ \\
$7$ & $0.3$ & $1.836993$ & $5.275083$ & $0.1884196$ & $-0.214972(2)$ \\
$7$ & $0.4$ & $1.653074$ & $5.247708$ & $0.1903568$ & $-0.213936(1)$ \\
$7$ & $0.45$ & $1.512406$ & $5.329219$ & $0.1949564$ & $-0.2183399(9)$ \\
$7$ & $0.49$ & $1.295776$ & $5.724418$ & $0.2103972$ & $-0.2397373(6)$ \\
$7$ & $0.499$ & $1.095652$ & $6.319841$ & $0.2317371$ & $-0.2745003(3)$ \\
$7$ & $0.4999$ & $0.9514141$ & $6.789595$ & $0.2483215$ & $-0.30448660(3)$ \\
$8$ & $0.3$ & $2.019755$ & $4.111256$ & $0.1507862$ & $-0.163772(1)$ \\
$8$ & $0.4$ & $1.867138$ & $3.857336$ & $0.1444159$ & $-0.152580(1)$ \\
$8$ & $0.5$ & $1.664691$ & $3.511329$ & $0.1352542$ & $-0.1377441(7)$ \\
$9$ & $0.1$ & $2.116662$ & $3.669228$ & $0.1352196$ & $-0.147865(1)$ \\
$9$ & $0.2$ & $2.051228$ & $3.564791$ & $0.1325754$ & $-0.1433588(9)$ \\
$9$ & $0.3$ & $1.940808$ & $3.387357$ & $0.1279803$ & $-0.1358046(8)$ \\
$9$ & $0.4$ & $1.783443$ & $3.131885$ & $0.1211342$ & $-0.1251445(7)$ \\
$9$ & $0.5$ & $1.576685$ & $2.791538$ & $0.1115804$ & $-0.1113160(5)$ \\
$10$ & $0.1$ & $1.978414$ & $3.129615$ & $0.1186709$ & $-0.1277554(7)$ \\
$10$ & $0.2$ & $1.913377$ & $3.031019$ & $0.1160724$ & $-0.1236493(7)$ \\
$10$ & $0.3$ & $1.804093$ & $2.864706$ & $0.1115965$ & $-0.1168034(6)$ \\
$10$ & $0.4$ & $1.649391$ & $2.627916$ & $0.1050164$ & $-0.1072221(5)$ \\
$10$ & $0.5$ & $1.448070$ & $2.317390$ & $0.09599833$ & $-0.0949289(4)$ \\
$15$ & $0.1$ & $1.316635$ & $1.699926$ & $0.07389641$ & $-0.0768709(1)$ \\
$15$ & $0.2$ & $1.266627$ & $1.635815$ & $0.07196285$ & $-0.0743140(1)$ \\
$15$ & $0.3$ & $1.183613$ & $1.529315$ & $0.06868680$ & $-0.0700771(1)$ \\
$15$ & $0.4$ & $1.068277$ & $1.381192$ & $0.06398800$ & $-0.0641991(1)$ \\
$15$ & $0.5$ & $0.9220001$ & $1.193071$ & $0.05775499$ & $-0.0567390(1)$ \\
$20$ & $0.1$ & $0.9230304$ & $1.103275$ & $0.05374213$ & $-0.05522177(4)$ \\
$20$ & $0.2$ & $0.8860353$ & $1.059177$ & $0.05226239$ & $-0.05340866(4)$ \\
$20$ & $0.3$ & $0.8249337$ & $0.9863242$ & $0.04976861$ & $-0.05040388(4)$ \\
$20$ & $0.4$ & $0.7406898$ & $0.8858366$ & $0.04621998$ & $-0.04623383(4)$ \\
$20$ & $0.5$ & $0.6349478$ & $0.7596359$ & $0.04156092$ & $-0.04093697(3)$ \\
\hline\hline
\end{tabular}
\caption{GSF correction to the generalized redshift invariant $\hatU$: sample numerical results.
Each row in the table corresponds to a particular geodesic orbit with parameters $(p,e)$ and associated frequencies $(\Omega_{r0},\Omega_{\varphi 0})$, as specified. The 5th column gives the numerical value of the coefficient $\alpha(p,e)$ [see Eq.\ (\ref{alpha})] needed for converting the Lorenz-gauge time coordinate $t$ to the ``asymptotically flat'' time $\hat t$. The sixth column displays the numerical values of the gauge-invariant quantity $\Delta\hatU$, computed via Eq.\ (\ref{DeltaUfinal}). Parenthetical figures indicate the estimated uncertainty in the last displayed decimals. 
}
\label{table:U_data}
\end{table}

In selecting the dataset for display in Table \ref{table:U_data} we have deliberately avoided parameter-space points which are located on or near the singular curve $p=p_s(e)$, where the quantity $\Delta\hatU$ becomes singular. This indeterminacy is an unavoidable price to pay for using the gauge-invariant parametrization ($\hOmegar,\hOmegaf$), which is ill-suited along $p_s(e)$.

It is worth clarifying a potentially confusing point regrading the interpretation of the data in Table \ref{table:U_data}. The table labels the orbits by their $(p,e)$-values and not by their invariant frequencies ($\hOmegar,\hOmegaf$) or the invariant parameters $(\pI,\eI)$ associated with them. This does {\em not} mean that we are measuring the GSF effect with respect to the background quantity $\hatUbg(p,e)$: we remind that the GSF correction $\Delta\hatU$ is defined in an invariant way with respect to $\hatUbg(\pI,\eI)$---recall Eq.\ (\ref{DeltaU}). The $p,e$ parametrization is adopted {\it a posteriori} for convenience, and it by no means compromises the gauge invariance of $\Delta\hatU$.

Finally, we should comment on how the information in Table \ref{table:U_data} could be used for comparisons with PN expressions, the latter being usually given in terms of the two invariant frequencies $\hOmegar,\hOmegaf$ (or simple combinations thereof; see, e.g., \cite{Arun:2007sg}) and not in terms of $p,e$. Suppose one has a PN expression for 
$\Delta\hatU$, given as a function of $\hOmegar,\hOmegaf$. One starts by extracting from the PN expression all terms through $O(\mu)$ (holding $\hOmegar,\hOmegaf$ fixed). One then subtracts the ``background'' [$O(\mu^0)$] quantity $\hatUbg(\hOmegar,\hOmegaf)$, where, of course, $\hOmegar,\hOmegaf$ are the full, perturbed frequencies. The remaining [$O(\mu)$] terms are explicit functions of $\hOmegar,\hOmegaf$, which may now be replaced with the background values $\Omega_{r0},\Omega_{\varphi0}$ [with the error from this replacement being of only  $O(\mu^2)$]. The resulting expression is readily evaluated for any given values of $p,e$ from Table \ref{table:U_data}, and the result may be compared with the corresponding GSF value of $\Delta\hatU$ given in the table. This procedure has been used to facilitate the comparison to be presented in Ref.\ \cite{inprep}.

\section{The circular-orbit limit and a tentative comparison with Numerical Relativity}\label{Sec:circ}

The limit $e\to 0$ (with fixed $p$) defines a circular orbit of radius $pM=r_\circ(={\rm const})$, which, however, still has two distinct frequencies associated with it: the azimuthal frequency $\Omega_\varphi$ (or $\hat\Omega_\varphi$) is simply $d\varphi/dt$ (or $d\varphi/d\hat t$) along the limiting circular orbit, and the radial frequency $\Omega_r$ (or $\hat\Omega_r$) is that associated with a slightly eccentric orbit resulting from an infinitesimal $e$-perturbation of the circular orbit. Hence, rather conveniently for us here, the circular-orbit limit defines a one-parameter family of orbits that are nonetheless characterized by two gauge-invariant frequencies. We may utilize the relation between these two frequencies (or between any two independent combinations thereof) as a simple gauge-invariant function for GSF studies. In Ref.\ \cite{Barack:2010ny} we have already studied the circular-orbit limit, and extracted the $O(\mu)$ gauge-invariant information embedded in the relation between the two frequencies (then used it to inform a comparison with PN-calibrated EOB models). 

In a recent numerical relativistic (NR) study \cite{Mroue:2010re} Mrou\'{e} {\it et al.}~report measurements of the periastron advance in fully nonlinear simulations of slightly eccentric black hole inspirals. Data are provided for a sample of binary mass ratios in the range 1:1--1:6, with eccentricities of order a few $\times 10^{-5}$ (see Fig.\ 8 of \cite{Mroue:2010re}; how the eccentricity is defined in these simulations is discussed in Sec.\ II therein).
It is interesting to examine the NR data against the predictions of our GSF calculation in the circular-orbit limit. Of course, we must exercise great caution in attempting such a comparison, and we note here three obvious caveats. First, even the ``small'' mass ratio of 1:6 is well outside the natural domain of validity of our $O(\mu)$ GSF approximation. Second, the NR simulation automatically accounts for dissipative effects, which are ignored in our GSF calculation. Our argument (Sec.\ \ref{subsec:diss}) that dissipation has a negligible effect on the periastron advance loses its validity when the mass ratio is not sufficiently small. Third, Ref.\ \cite{Mroue:2010re} makes no special effort to obtain very accurate precession data, since its main motivation lies somewhere else (it attempts to develop a method for eliminating spurious eccentricity in quasicircular inspirals). As a result, the statistical variance (due to numerical error) in the precession data of \cite{Mroue:2010re} is large, making a meaningful comparison difficult. The numerical error is particularly large for smaller mass ratios, which are, alas, most useful to us here. 

Nonetheless, we would like here to take advantage of this opportunity to make a first contact between the GSF and NR programs, if only to point to the potential of a mutually beneficial synergy between the two programs, and in order to motive further study. 

Mrou\'{e} {\it et al.}~plot the frequency ratio $\hOmegaf/\hOmegar$ versus the adimensionalized azimuthal frequency
$(M+\mu)\hOmegaf$. Note we must interpret the NR frequencies as our ``hat'' frequencies (those defined with respect to the ``asymptotically flat'' time $\hat t$) rather than our ``Lorenz-gauge'' frequencies $\Omegaf,\Omegar$. It is also crucial to notice that Mrou\'{e} {\it et al.} adimensionalize the frequency using the {\em total} mass $M+\mu$ and not (as customary in GSF analysis) the large mass $M$. A GSF expression for $\hOmegaf/\hOmegar$ is readily obtained, through $O(\mu)$, using Eqs.\ (3), (6), and (14) of Ref.\ \cite{Barack:2010ny}:
\begin{eqnarray}\label{freqratio}
\frac{\hOmegaf}{\hOmegar}=\frac{\Omegaf}{\Omegar} 
=
\left( 1-\frac{6M}{r_\circ} \right)^{-3/2} \left[
1 - \frac{6M}{r_\circ}
+ \frac{r_\circ(r_\circ-3M)}{\mu M}F_\circ^r
- \frac{r_\circ(r_\circ-3M)}{2 \mu M} F_1^r
+ \frac{(r_\circ-3M)^{3/2}}{\mu M^{1/2} r_\circ^2} F^1_\varphi
\right].
\end{eqnarray}
Here $r_\circ$ is the radius of the limiting circular orbit, and the $O(\mu^2)$ quantities $F_\circ^r$, $F_1^r$ and $F^1_\varphi$ arise from the formal $e$-expansion of the conservative GSF components along the slightly eccentric orbit. Specifically, $F_\circ^r$ is the $r$ component of the conservative GSF along the strictly circular orbit, and $F_1^r,F^1_\varphi$ are associated with the $O(e)$ variation of the GSF. We refer the reader to \cite{Barack:2010ny} for a precise definition of these GSF quantities. How these quantities are extracted in practice from the numerical GSF data is described in Ref.\ \cite{Barack:2010tm}. The azimuthal frequency is given through $O(\mu)$ by 
\begin{equation}
\hOmegaf =
\sqrt{\frac{M}{r_\circ^3}}
\left[ 1 - \alpha_\circ
- \frac{r_\circ^2(r_\circ-3M)}{2\mu M(r_\circ-2M)} F_\circ^r\right]
\end{equation}
[see Eqs.\ (3) and (6) of \cite{Barack:2010ny}], where $\alpha_\circ\equiv \alpha(e=0)=\mu\left[r_0(r_0-3M)\right]^{-1/2}$. Hence, through $O(\mu)$,
\begin{equation}\label{hatOmegaf}
(M+\mu)\hOmegaf =
(M/r_\circ)^{3/2}
\left[ 1 +\mu - \alpha_\circ
- \frac{r_\circ^2(r_\circ-3M)}{2\mu M(r_\circ-2M)} F_\circ^r\right].
\end{equation}

\begin{figure}[htb]
\includegraphics[height=8cm]{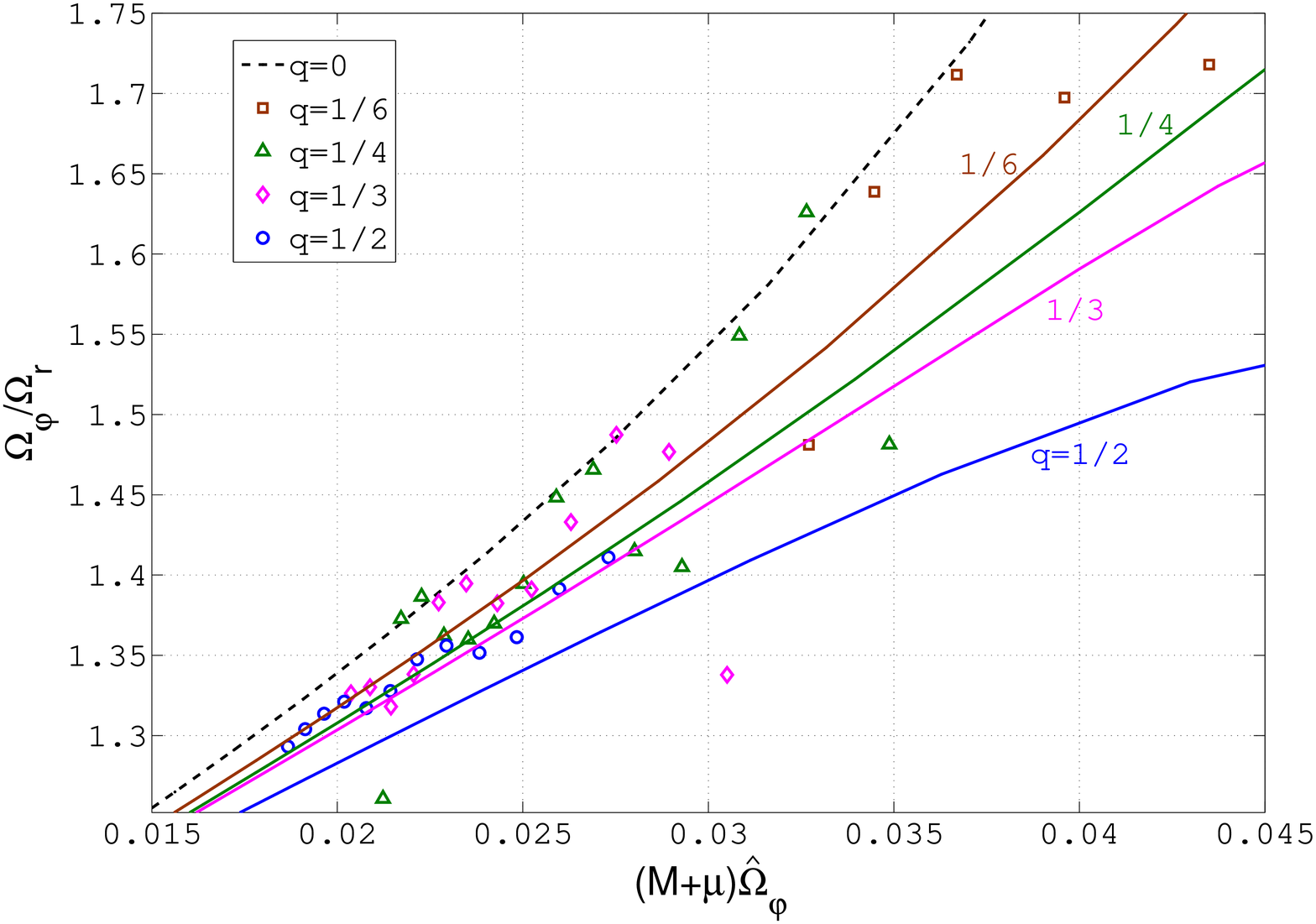}
\caption{(color online) Tentative comparison of GSF and NR data for the periastron advance of slightly eccentric orbits. Shown is the frequency ratio $\Omega_\varphi/\Omega_r$ as a function of $\hOmegaf$ (adimensionalized using the total mass $M+\mu$), for a variety of mass ratios $q=\mu/M$ between 1:2 and 1:6. The dashed line corresponds to a test particle ($q=0$). Single data points describe results from NR simulations, reproduced here from Fig.\ 8 of Mrou\'{e} {\it et al.} \cite{Mroue:2010re}. Solid lines are interpolated $O(q)$ GSF ``predictions'', calculated using Eqs.\ (\ref{freqratio}) and (\ref{hatOmegaf}) with the numerical values of the GSF coefficients tabulated in Appendix \ref{App:circ}. The horizontal scale of this plot roughly coincides with that of Fig.\ 8 of \cite{Mroue:2010re} for easy reference. Despite the manifest low accuracy of the NR data, this preliminary comparison is already rather instructive (as described in the text), and motivates further study.
}
\label{fig:NR}
\end{figure}

In Appendix \ref{App:circ} we give numerical values for the GSF coefficients $F_\circ^r(r_\circ)$, $F_1^r(r_\circ)$ and $F^1_\varphi(r_\circ)$. Using these values in Eqs.\ (\ref{freqratio}) and (\ref{hatOmegaf}) we can obtain a numerical (parametric) relation between $\hOmegaf/\hOmegar(=\Omegaf/\Omegar)$ and $(M+\mu)\hOmegaf$ for any given value of $\mu$. We plot this relation in Fig.\ \ref{fig:NR} for a sample of mass ratios $q\equiv\mu/M$. Superposed on these GSF ``predictions'' (which, of course, represent gross extrapolations beyond the natural mass-ratio domain of the GSF) we display the NR data points from Mrou\'{e} {\it et al.}~\cite{Mroue:2010re}. Each NR data set, for a given $q$, comes from a single inspiral simulation; how the frequencies are extracted from the numerical data is explained in Ref.\ \cite{Mroue:2010re}.

The NR data clearly resolve a nonzero precession effect, but they are not accurate enough to allow a detailed quantitative comparison with the GSF predictions. Nevertheless, one can make several tentative observations.
(i) The NR and GSF data are in agreement on the ``sign'' of the post-geodesic precession effect: it is opposite that of the geodesic precession, i.e., the GSF acts to {\em reduce} the rate of periastron advance. (ii) The NR data for $q=1/2$ is least noisy and perhaps most accurate. If we are to trust these data, we find that the GSF prediction ``overestimates'' the post-geodesic effect by about a factor 2. This is not unreasonable for a mass ratio as large as 1:2. We see how our comparison already starts to tell us about the sign and magnitude of uncalculated higher-order GSF contributions. Tentatively, it would seem that the 2nd-order GSF precession effect is opposite in sense to that of the 1st-order effect. (iii) A most meaningful comparison (relatively speaking) would have been provided by the 1:6 data, if not for the very large scatter of the NR points in this case. The 1:6 NR data seem roughly evenly distributed about the GSF curve, but it is not possible to make more definite statements. 

The above tentative comparison illustrates the potential benefits from synergic GSF/NR studies. Foremostly, it provides a strong two-way test of the results, because GSF and NR computations use highly independent methods. From the point of view of NR practitioners, the GSF predictions provide an accurate benchmark against which to assess the quality of the numerical simulations. From the GSF point of view, comparison with NR simulations gives access to valuable information about the effect of currently inaccessible high-order GSF corrections. We envisage a synergy between GSF and NR methods as a fast-track avenue to the modelling of the two-body dynamics in intermediate mass-ratio inspirals, which is currently beyond the reach of either method.   

Motivated by the above, we have recently initiated a collaborative study to pursue and exploit a more detailed comparison of the GSF and NR precession data. Preliminary new NR data by Mrou\'{e} {\it et al.}~are dramatically more accurate, and show a remarkable agreement with the GSF predictions at $q=1/8$. We hope to report results from this study in a forthcoming paper \cite{NRvsGSF}.

\section*{ACKNOWLEDGEMENTS}
We are grateful to Thibault Damour and Amos Ori for many discussions that influenced the development of this work. We thank Alexandre Le Tiec for helping us test the results of Sec.\ \ref{Sec:U} against PN expressions, and to Abdul Mrou\'{e} for providing the NR data for Fig.\ \ref{fig:NR}.
LB acknowledges support from STFC through Grant No.~PP/E001025/1.
NS acknowledges supports by the Grant-in-Aid for Scientific Research
 (No.~10015318) and the Global COE Program ``The Next Generation of
 Physics, Spun from Universality and Emergence'', from the Ministry of
 Education, Culture, Sports, Science and Technology of Japan.


\appendix

\section{Isofrequency geodesic orbits}\label{App:sing}

We think there is a common belief that bound geodesic orbits in Schwarzschild (or Kerr) spacetime can be labelled uniquely by their frequencies. This belief is unfounded, and turns out to be false. To the best of our knowledge, this issue was (rather surprisingly) never addressed in the literature, so we briefly discuss it here. 

\begin{figure}[htb]
\includegraphics[height=9cm]{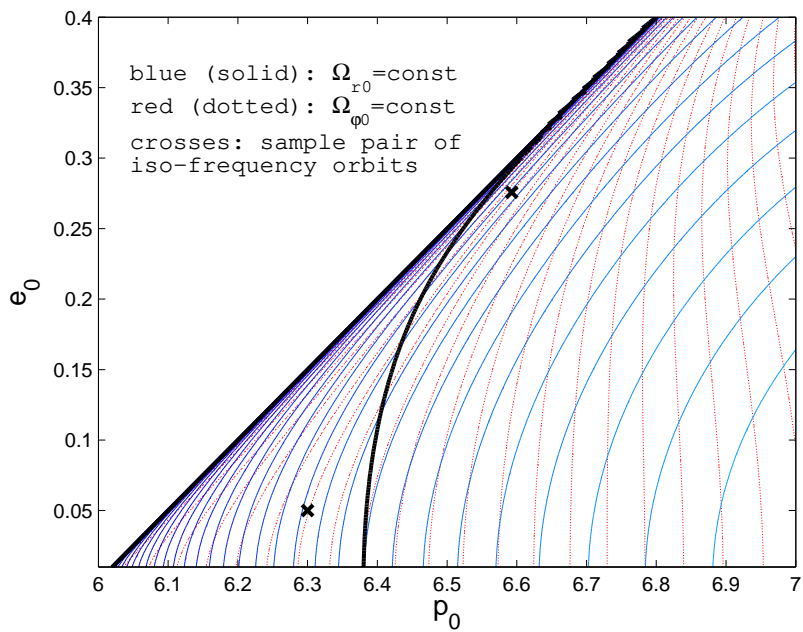}
\caption{(color online) Singularity of the transformation $(p_0,e_0)\to (\Omega_{\varphi0},\Omega_{r0})$ for bound (eccentric) geodesics in Schwarzschild spacetime. The plot displays the level lines $\Omega_{\varphi0}$=const (dotted, red) and $\Omega_{r0}$=const (solid, blue) over a portion of the $p_0,e_0$ parameter space. The diagonal $p_0=6+2e_0$ (straight black line) is the separatrix, where $\Omega_{r0}=0$; stable geodesic orbits exist in the region $p_0>6+2e_0$. In the domain shown, $\Omega_{\varphi0}$ decreases monotonically with $p_0$ while $\Omega_{r0}$ {\em increases} monotonically with $p_0$ (the sign of $\partial\Omega_{r0}/\partial p_0$ reverses further out to the right of the region shown). The thick (black) curve, $p_0=p_s(e_0)$, is the locus of points where the $\Omega_{\varphi0}$ and $\Omega_{r0}$ level lines are tangent to one another; along this curve the Jacobian of the transformation $(p_0,e_0)\to (\Omega_{\varphi0},\Omega_{r0})$ vanishes. Each orbit left of the singular curve has a ``dual'' isofrequency orbit associated with it, which lies to the right of the singular curve. One such pair is indicated in the plot (black crosses), at $(p_0,e_0)=(6.3,0.05)$ and $(6.59274,0.27569)$.
}
\label{fig:sing}
\end{figure}

In Fig.\ \ref{fig:sing} we plot the level lines $\Omega_{\varphi0}$=const and $\Omega_{r0}$=const over a portion of the $p_0,e_0$ parameter space of bound geodesics in Schwarzschild. Recall the region of parameter space with $p_0>6+2e_0$ (and $0\leq e_0<1$) corresponds to bound geodesics. The transformation $(p_0,e_0)\to(\Omega_{r0},\Omega_{\varphi0})$ becomes singular along the separatrix $p_0=6+2e_0$. More surprisingly, it also turns out to be singular along a certain curve $p_0=p_s(e_0)$, well outside the separatrix, which divides the parameter space into two disjoint domains, $p_0<p_s(e_0)$ and $p_0>p_s(e_0)$ (see Fig.\ \ref{fig:sing}). At each point along $p_s(e_0)$, the $\Omega_{\varphi0}$=const level line is tangent to an $\Omega_{r0}$=const level line, and the Jacobian matrix of the transformation $(p_0,e_0)\to(\Omega_{r0},\Omega_{\varphi0})$ becomes singular. Note that the transformation $(E_0,L_0)\to(p_0,e_0)$ is perfectly regular across the entire parameter space of bound geodesics, so one cannot dismiss the above behavior as a mere peculiarity of the $p_0,e_0$ parametrization.

It can be easily verified that the presence of the singular curve $p_0=p_s(e_0)$ results in the following: for each orbit belonging to the ``left'' domain [$p_0<p_s(e_0)$], there exists a ``dual'' (or ``isofrequency'') orbit in the ``right'' domain [$p_0>p_s(e_0)$], which is physically distinct but has the same frequencies $\Omega_{\varphi0},\Omega_{r0}$.  For example, the orbit with parameters $(p_0,e_0)=(6.3,0.05)$ has the same frequencies as an orbit with parameters $(p_0,e_0)=(6.59274\ldots,0.27569\ldots)$. This, of course, means that the frequency pair ($\Omega_{\varphi0},\Omega_{r0}$) does not label the geodesic orbits uniquely. However, one may still use the ($\Omega_{\varphi0},\Omega_{r0}$) parametrization separately in each of the ``left'' and ``right'' domains (as we do in Sec.\ \ref{Sec:U}). 

It should be noted that all orbits on the ``left'' domain are deep within the zoom-whirl regime. The ``least'' zoom-whirling orbit in the left domain [the one corresponding to the intersection of $p_s(e_0)$ with $e_0=0$] has a frequency ratio $\Omega_{\varphi0}/\Omega_{r0}\simeq 4.1$, i.e., it executes over 3 whirls per radial period. It is also interesting to ask whether the range of dual orbits in the ``right'' domain extends to the weak-field regime (which would suggest fascinating astrophysical implications). It is clear, however, that this is not the case: the azimuthal frequencies of orbits in the left domain (and hence those of their ``right'' duals too) are confined to the range  $6.38^{-3/2}\lesssim M\Omega_{\varphi0}< 4^{-3/2}$.

Readers familiar with the literature on radiation reaction in black hole spacetimes may find the curve $p_s(e_0)$ in Fig.\ \ref{fig:sing} here reminiscent of the ``critical curve'' shown in Fig.\ 3 of Cutler {\it et al.}~\cite{Cutler:1994pb}, which describes the locus of points in the $p_0,e_0$ plane at which the radiative evolution of the eccentricity changes its sense (so that $de/dp$ turns from negative to positive). Inspection of the two plots reveals, however, that the two curves lie sufficiently far apart in the parameter space to assume that there is no direct relation between them. It is still interesting to ask about possible anomalies in the behavior of a radiatively inspiraling object as it crosses the curve $p_s(e_0)$. If the radiative evolution happens to drive the inspiral across $p_s(e_0)$ in a direction (in the $p_0,e_0$ plane) tangent, or nearly tangent, to the frequency level lines, we might expect a ``hang up'' episode during which the evolution of the two frequencies halts, or slows down. The information in Fig.\ 3 of Ref.\ \cite{Cutler:1994pb} suggests, however, that the radiative evolution through $p_s(e_0)$ proceeds in a direction roughly {\em orthogonal} to the frequency level lines. Any ``hang up'' effect should therefore be minimal.

One may also ponder the possibility of finding resonant interaction effects acting between pairs of dual orbits around astrophysical black holes (e.g., between clumps of accreting matter). This intriguing possibility deserves exploration. A preliminary step would be to understand the dual behavior across the full (3-dimensional) parameter space of generic orbits in Kerr geometry. We have confirmed that a similar duality exists at least in the subspace of eccentric equatorial geodesics in Kerr \cite{Niels}.

Finally, we point to the fact that, in the context of the GSF problem, the relation between the two frequencies along the (GSF-perturbed) singular curve, $\Omega_\varphi=\Omega_{\varphi s}(\Omega_r)$, is a {\em gauge-invariant} one. As such, we can envisage it being utilized as a reference for comparison between different calculations of the GSF, and for a strong-field calibration of approximate analytic methods---in much the same way that the relation $\Omega_{\varphi}(\Omega_r)$ has been utilized in the circular-orbit limit \cite{Damour:2009sm,Barack:2010ny}. We hope to explore this possibility in future work.

\section{Treatment of numerical $\chi$-integrals near periastron and apastron} \label{AppA}

The numerical implementation of formulas (\ref{Deltadelta}) (for $\Delta\delta$), (\ref{DeltaT}) (for $\Delta T$) and  (\ref{DeltaTau}) (for $\Delta\tilde\calT$) is somewhat subtle, due to the formal divergence of individual terms in the respective integrands $\Delta\varphi_{\chi}$, $\Delta t_{\chi}$ and $\Delta\tau_{\chi}$ at the two radial turning points, $\chi=0,\pi$. As we explained in the text, each of the full integrands is in fact smooth for all $\chi$, including at $\chi=0,\pi$. However, in practice, the divergence of individual terms require a special treatment at the turning points, which we describe here. In what follows we refer specifically to $\Delta\delta$ (and $\Delta\varphi_{\chi}$) for concreteness; $\Delta T$ and $\Delta\tilde\calT$ are treated in a similar manner.

Consider first the periapsis, $\chi=0$, which is more easily dealt with. The function $\Delta\varphi_{\chi}(\chi)$ [Eq.\ (\ref{Deltaomega2})] has the form 
\begin{equation}\label{form1}
\Delta\varphi_{\chi}=f_1(\chi)+f_2(\chi)\frac{{\cal E}(\chi)}{\sin^2\chi}+f_3(\chi)\frac{{\cal L}(\chi)}{\sin^2\chi},
\end{equation}
where $f_1,f_2,f_3$ are certain functions of $\chi$ which are regular (smooth) at $\chi=0$ ($\Delta t_{\chi}$ and $\Delta\tau_{\chi}$ have similar forms, with different $f_n$'s). The functions ${\cal E}(\chi)$ and ${\cal L}(\chi)$ have even Taylor expansions at $\chi=0$, with ${\cal E}(0)={\cal L}(0)=0$, and so the expression in Eq.\ ({\ref{form1}) is in fact perfectly regular at $\chi=0$. Still, the factors ${\cal E}/\sin^2\chi$ and ${\cal L}/\sin^2\chi$ pose a practical problem, because the numerical integration routine we apply to evaluate Eqs.\ (\ref{Deltadelta}) requires as input (also) the value $\Delta\varphi_{\chi}(0)$. We deal with this simply by writing 
\begin{eqnarray}\label{limit1}
\lim_{\chi\rightarrow 0} \frac{{\cal E}(\chi)}{\sin^2\chi}
&=& \frac{1}{2}\left.\frac{d^2\cal E}{d\chi^2}\right|_{\chi=0}
=
- \left.\frac{1}{2\mu}\left(
\frac{dF_t^{\rm cons}}{d\chi} \frac{d\tau}{d\chi}
\right)\right|_{\chi=0}, 
\nonumber\\
\lim_{\chi\rightarrow 0} \frac{{\cal L}(\chi)}{\sin^2\chi}
&=& \frac{1}{2}\left.\frac{d^2\cal L}{d\chi^2}\right|_{\chi=0}
=
\left.\frac{1}{2\mu}\left(
\frac{dF_\varphi^{\rm cons}}{d\chi} \frac{d\tau}{d\chi}
\right)\right|_{\chi=0},
\end{eqnarray}
where we have recalled the definitions of ${\cal E}$ and ${\cal L}$ in Eqs.\ (\ref{calEL}). The factor $d\tau/d\chi$ is given explicitly in Eq.\ (\ref{dtaudchi}), and in the above equalities we have made use of the fact that it is an even function of $\chi$, so the term $\propto d^2\tau/d\chi^2$ vanishes at $\chi=0$. The GSF derivatives in Eqs.\ (\ref{limit1}) are readily evaluated numerically at $\chi=0$ from our GSF data sets. Then the numerical value of $\Delta\varphi_{\chi}(0)$ is obtained by substituting from Eqs.\ (\ref{limit1}) in the form (\ref{form1}). 

Now turn to consider the apastron, $\chi=\pi$, where the situation is slightly more involved. For this discussion, we note that $\Delta\varphi_{\chi}$ can also be written in the form 
\begin{equation}\label{form2}
\Delta\varphi_{\chi}=\left[\frac{g_1(\chi){\cal E}(\pi)-g_2(\chi){\cal E}(\chi)}{\sin^2\chi}\right]
+
\left[\frac{g_3(\chi){\cal L}(\pi)-g_4(\chi){\cal L}(\chi)}{\sin^2\chi}\right],
\end{equation}
where the $g_n$'s are certain functions of $\chi$ which are smooth near $\chi=\pi$, where they satisfy 
\begin{equation}
g_1(\chi)=g_2(\chi)+O(\chi-\pi)^2,
\quad\quad
g_3(\chi)=g_4(\chi)+O(\chi-\pi)^2.
\end{equation}
($\Delta t_{\chi}$ can be written in the same form, with different functions $g_n$ having the same properties. The same also applies to $\Delta\tau_{\chi}$, modulo an additive function of $\chi$ which is however smooth at $\chi=\pi$ and hence of no concern to us here.)
Since $\cal E$ and $\cal L$ admit regular Taylor expansions about $\chi=\pi$, of the form ${\cal E}={\cal E}(\pi)+O(\chi-\pi)^2$ and ${\cal L}={\cal L}(\pi)+O(\chi-\pi)^2$, it follows that the expression in Eq.\ (\ref{form2}) is perfectly regular at $\chi=\pi$, and, in particular, the limit $\chi\to\pi$ of this expressions is finite. Here, however, it is not sufficient to obtain $\Delta\varphi_{\chi}(\pi)$ via a Taylor expansion as we did  for $\chi=0$.  An added practical difficulty is that, for values of $\chi$ near $\pi$, a delicate cancellation occurs in Eq.\ (\ref{form2}) between the $g_1$ and $g_2$ terms and also between the $g_3$ and $g_4$ terms. This can give rise to large numerical errors due to roundoff if one attempts to evaluate Eq.\ (\ref{form2}) directly. 

To circumvent this problem, we introduce 
\begin{eqnarray}
\bar{\cal E}(\chi) &\equiv &
-\int_{\chi}^{\pi}F_t^{\rm cons}(\chi')
\frac{d\tau}{d\chi'}d\chi' = {\cal E}(\pi)-{\cal E}(\chi), 
\nonumber\\
\bar{\cal L}(\chi) &\equiv &
\int_{\chi}^{\pi}F_\varphi^{\rm cons}(\chi')
\frac{d\tau}{d\chi'}d\chi' = {\cal L}(\pi)-{\cal L}(\chi),
\end{eqnarray}
in terms of which Eq.\ (\ref{form2}) becomes
\begin{equation}\label{form3}
\Delta\varphi_{\chi}=h_1(\chi){\cal E}(\pi)+g_2(\chi)\frac{\bar{\cal E}(\chi)}{\sin^2\chi}
+h_2(\chi){\cal L}(\pi)+g_4(\chi)\frac{\bar{\cal L}(\chi)}{\sin^2\chi}
\equiv \bar \Delta\varphi_{\chi},
\end{equation}
where $h_1(\chi)\equiv [g_1(\chi)-g_2(\chi)]/\sin^2\chi$ and $h_2(\chi)\equiv [g_3(\chi)-g_4(\chi)]/\sin^2\chi$ are regular (smooth) at $\chi=\pi$.  This form no longer involves a delicate cancellation between different terms near $\chi=\pi$ and is thus free from the above numerical difficulty. To obtain $\Delta\varphi_{\chi}(\pi)$ we simply use
\begin{eqnarray}\label{limit2}
\lim_{\chi\rightarrow \pi} \frac{\bar{\cal E}(\chi)}{\sin^2\chi}
&=& \frac{1}{2}\left.\frac{d^2\bar{\cal E}}{d\chi^2}\right|_{\chi=\pi}
=
\left.\frac{1}{2\mu}\left(
\frac{dF_t^{\rm cons}}{d\chi} \frac{d\tau}{d\chi}
\right)\right|_{\chi=\pi},
\nonumber\\
\lim_{\chi\rightarrow \pi} \frac{\bar{\cal L}(\chi)}{\sin^2\chi}
&=& \frac{1}{2}\left.\frac{d^2\bar{\cal L}}{d\chi^2}\right|_{\chi=\pi}
=
- \left.\frac{1}{2\mu}\left(
\frac{dF_\varphi^{\rm cons}}{d\chi} \frac{d\tau}{d\chi}
\right)\right|_{\chi=\pi},
\end{eqnarray}
where the GSF derivatives are evaluated from the numerical data.

Finally, to carry out the integral in Eq.\ (\ref{Deltadelta}), we split the integration domain as
\begin{equation}
\Delta\delta= \frac{1}{\pi}\int_0^{\pi/2}\Delta\varphi_{\chi} d\chi
+ \frac{1}{\pi} \int_{\pi/2}^{\pi}\bar\Delta\varphi_{\chi} d\chi,
\end{equation}
where we use the form (\ref{form1}) [with (\ref{limit1})] for the first integral, and the form (\ref{form3}) [with (\ref{limit2})] for the second. The integrals for $\Delta T$ [Eq.\ (\ref{DeltaT})] and for $\Delta\tilde\calT$  [Eq.\ (\ref{DeltaTau})] are evaluated in the same way. 

\section{Relation between ``true'' and ``conservative-only'' periastron advance}\label{AppB}

In this appendix we establish the relation (\ref{true}), which states that the periastron advance $\delta_{\rm true}$ of the physical, evolving orbit [as it is defined in Eq.\ (\ref{delta_true})] is equal through $O(\mu)$ to the conservative advance $\delta$ associated with a certain ``average'' conservative orbit. 

We begin by reproducing Eqs.\ (\ref{Omegaf_true}) and  (\ref{delta_true}) here for easy reference:
\begin{equation}\label{delta_true2}
\delta_{\rm true}=(2\pi)^{-1}\Phi_{\rm true}  - 1, \quad\quad
\Phi_{\rm true}=\int_{\tilde\chi_1}^{\tilde\chi_2}\tilde\varphi_{\chi}[\tilde\chi;\tilde p(t(\tilde\chi)),\tilde e(t(\tilde\chi))] d\tilde\chi.
\end{equation}
Here, recall, the integration is over a complete radial cycle of the slowly evolving orbit, from an apastron at $\tilde\chi=\tilde\chi_1$ ($t=t_1$) to the next apastron at $\tilde\chi=\tilde\chi_2$ ($t=t_2$). The values of $\tilde\chi_1$ and $\tilde\chi_2$ are found from the conditions $\tilde\chi_1\to -\pi$ and $\tilde\chi_2\to \pi$ for $\mu\to 0$, together with ${\tilde r}'(\tilde\chi_1)=\tilde r'(\tilde\chi_2)=0$, where throughout this appendix a prime denotes $d/d\tilde\chi$. Using Eq.\ (\ref{hatrofchi}), these conditions give
\begin{eqnarray}\label{hatchi12}
\tilde\chi_1 &=& -\pi + \delta\chi(p_1,e_1) + O(\mu^2), \nonumber\\
\tilde\chi_2 &=&  \pi + \delta\chi(p_2,e_2) + O(\mu^2),
\end{eqnarray}
with
\begin{equation}
\delta\chi(p,e) = \frac{p'(1-e)+pe'}{pe}.
\end{equation}
Here it is sufficient to evaluate $p'$ and $e'$ at leading order [$O(\mu)$], which may be done with the help of Eqs.\ (\ref{hatp}) and (\ref{dtdchi}). Recall $(p_1,e_1)$ and $(p_2,e_2)$ are the orbital parameters at times $t_1$ and $t_2$, respectively (assumed given). Since $\delta\chi$ is already $O(\mu)$, in Eq.\ (\ref{hatchi12}) we are allowed to replace $p_1,p_2\to \bar p$ and $e_1,e_2\to \bar e$, where $\bar p,\bar e$ are the average parameter values defined in Eq.\ (\ref{av}). $\bar p,\bar e$ are also, through $O(\mu)$, the parameter values at the periastron. We thus have
\begin{eqnarray}\label{hatchi12b}
\tilde\chi_1 &=& -\pi + \delta\chi(\bar p,\bar e) + O(\mu^2), \nonumber\\
\tilde\chi_2 &=&  \pi + \delta\chi(\bar p,\bar e) + O(\mu^2).
\end{eqnarray}

Now let us formally expand $\tilde p$ and $\tilde e$ in $\tilde\chi$ about the periastron through $O(\mu)$, noting that the periastron value of $\tilde\chi$ is $O(\mu)$:
\begin{equation}
\tilde p = \bar{p} + p'(\bar{p},\bar{e})\tilde\chi + O(\mu^2),
\quad
\tilde e = \bar{e} + e'(\bar{p},\bar{e})\tilde\chi + O(\mu^2),
\label{eq:pe-expand}
\end{equation}
where $p',e'$ [each of $O(\mu)$] are evaluated at the periastron. We use this to expand the integrand in Eq.\ (\ref{delta_true2}) in the form
\begin{eqnarray}
\Phi_{\rm true}&=&
\int_{\tilde\chi_1}^{\tilde\chi_2}d\tilde\chi \left[
\tilde\varphi_{\chi}(\tilde\chi;\bar{p},\bar{e})
+ \left.\frac{\partial\tilde\varphi_{\chi}(\tilde\chi;\tilde{p},\tilde{e})}{\partial \tilde p}\right|_{\bar p,\bar e}p'\tilde\chi
+ \left.\frac{\partial\tilde\varphi_{\chi}(\tilde\chi;\tilde{p},\tilde{e})}{\partial \tilde e}\right|_{\bar p,\bar e}e'\tilde\chi
+ O(\mu^2)
\right]
\nonumber\\
&\equiv& \Phi_1+\Phi_2+\Phi_3,
\label{eq:fullT1}
\end{eqnarray}
where $\Phi_1$, $\Phi_2$ and $\Phi_3$ denote the corresponding contributions to $\Phi_{\rm true}$ from the first, second and third terms in the integrand [neglecting the terms of $O(\mu^2)$].

Let us consider $\Phi_2$ first.  Using Eq.\ (\ref{hatchi12b}) we can express it as  
\begin{equation}
\Phi_2=
\left(\left.\frac{\partial\tilde\varphi_{\chi}(\pi;\tilde{p},\tilde{e})}{\partial \tilde p}\right|_{\bar p,\bar e}
-\left.\frac{\partial\tilde\varphi_{\chi}(-\pi;\tilde{p},\tilde{e})}{\partial \tilde p}\right|_{\bar p,\bar e}
\right)p' \pi\delta\chi 
+
\int_{-\pi}^{\pi}d\tilde\chi
\left.\frac{\partial\tilde\varphi_{\chi}(\tilde\chi;\tilde{p},\tilde{e})}{\partial \tilde p}\right|_{\bar p,\bar e}p'\tilde\chi
+O(\mu^2).
\label{eq:fullT-diss1}
\end{equation}
The first two terms on the right-hand side may be evaluated at the geodesic limit, since they are multiplied by $\delta\chi\propto O(\mu)$. We note, recalling Eq.\ (\ref{dphidchi}), that $\varphi_{\chi}(\chi;p,e)$ is an even function of $\chi$, and so is its partial derivative with respect to $p$. Therefore, the first two terms on the right-hand side in Eq.\ (\ref{eq:fullT-diss1}) cancel each other through $O(\mu)$. In addition, we note that the integrand in the third term is odd in $\tilde\chi$ and it follows that the integral vanishes. Hence, the entire contribution $\Phi_2$ is $O(\mu^2)$, and similarly for $\Phi_3$:
\begin{equation}\label{Phi23}
\Phi_2,\Phi_3=O(\mu^2).
\end{equation}

Concentrate then on $\Phi_1$. Using Eq.\ (\ref{hatchi12b}) again we have
\begin{equation}
\Phi_1=
\left[\tilde\varphi_{\chi}(\pi;\bar{p},\bar{e})
- \tilde\varphi_{\chi}(-\pi;\bar{p},\bar{e})\right]\delta\chi
+\int_{-\pi}^{\pi}\tilde\varphi_{\chi}(\tilde\chi;\bar{p},\bar{e})d\tilde\chi
+ O(\mu^2) ,
\label{eq:fullT-cons}
\end{equation}
in which the first two terms cancel each other through $O(\mu)$ by virtue of the aforementioned even parity of $\tilde\varphi_{\chi}(\chi;p,e)$, and the integral is simply the quantity ${\Phi}(\bar{p},\bar{e})$ associated with the conservative orbit with parameters $\bar{p},\bar{e}$. Hence,
\begin{equation}\label{Phi1}
\Phi_1={\Phi}(\bar{p},\bar{e}) + O(\mu^2).
\end{equation}

Combining Eqs.\ (\ref{eq:fullT1}), (\ref{Phi23}) and (\ref{Phi1}) we conclude
\begin{equation}
\Phi_{\rm true} =\Phi(\bar{p},\bar{e}) + O(\mu^2),
\end{equation}
and substituting $\Phi_{\rm true}=2\pi(\delta_{\rm true}+1)$ and $\Phi(\bar p,\bar e)=2\pi[\delta(\bar p,\bar e)+1]$ leads directly to Eq.\ (\ref{true}).

\section{Mode-sum regularization of $H^R$}\label{App:H^R}

In this appendix we describe the calculation of the R-field combination 
\begin{equation}\label{HRdef}
H^R=\frac{1}{2}h_{\alpha\beta}^R u^\alpha u^\beta
\end{equation}
(evaluated along the orbit) that goes into the expression for our gauge invariant $\Delta\hatU$ [Eq.\ (\ref{DeltaUfinal}) with (\ref{DeltaTau}) in Sec.\ \ref{Sec:U}]. Our numerical code returns the tensor-harmonic modes of the {\em retarded} metric perturbation (not those of the $R$-field $h_{\alpha\beta}^R$), and the construction of the function $H^R$ follows through a certain regularization procedure, which resembles the standard mode-sum regularization of the GSF. Here we derive the necessary ``mode-sum regularization formula'' for the quantity $H^R$. Our analysis follows very closely the method of Refs.\ \cite{Barack:2001gx,Barack:2002mha,Barack:2002bt}, although the details are much simpler in our present case.

We start by recalling 
\begin{equation}\label{HR}
h_{\alpha\beta}^R = h^{\rm full}_{\alpha\beta} - h_{\alpha\beta}^S,
\end{equation}
where $h_{\alpha\beta}^{\rm full}$ is the full (retarded) metric  perturbation (denoted simply $h_{\alpha\beta}$ elsewhere in this work) and $h_{\alpha\beta}^S$ is Detweiler--Whiting's S-field \cite{Detweiler:2002mi}. Both $h_{\alpha\beta}^S$ and $h_{\alpha\beta}^R$ are uniquely defined (in terms of a particular Green function) in a small neighborhood of the particle's worldline, with the field $h_{\alpha\beta}^R$ being a smooth ($C^\infty$) homogeneous solution of the linearized field equations. At a field point $x\equiv x^{\alpha}$ in the vicinity of a point $z\equiv z^{\alpha}$ on the worldline, the S-field admits the local expansion \cite{Barack:2001gx}
\begin{equation}
\bar{h}_{\alpha\beta}^{S}(x) =
\frac{4\mu {v}_\alpha {v}_\beta}{\epsilon}
\left[ 1 + O(\delta x^2) \right],
\end{equation}
where an overbar denotes trace reversal, $v_{\alpha}(x;z)$ is the four-velocity vector parallel-propagated from $z$ to $x$ along a short geodesic section connecting the two points, $\epsilon(x;z)$ is the spatial geodesic distance from $x$ to the worldline, and $\delta x\equiv x-z$. The spatial distance $\epsilon$ itself has the local expansion 
\begin{equation}
\epsilon = \epsilon_0\left( 1 + \frac{S_1}{2\epsilon_0^2} \right)
+ O(\delta x^3),
\label{eq:exp-geod-dist}
\end{equation}
in which 
$\epsilon_0^2 =(g_{\alpha\beta}+u_{\alpha}u_{\beta})\delta x^\alpha \delta x^\beta$ (with the metric function and four velocity evaluated at $z$) and $S_1$ is a cubic polynomial in $\delta x$ [given explicitly in Eq.\ (A5) of \cite{Barack:2001gx}].
Defining now the field $H^S(x)\equiv\frac{1}{2}h_{\alpha\beta}^S v^\alpha v^\beta$, we find
\begin{equation}
H^S(x)=
\frac{1}{2}\left(
\bar{h}_{\alpha\beta}^{S}
- \frac{1}{2} \bar{h}^{S}_{\mu\nu}g^{\mu\nu} g_{\alpha\beta}
\right)  {v}^\alpha {v}^\beta
\nonumber=
\frac{\mu}{\epsilon} + O(\delta x),
\label{eq:huuS1}
\end{equation}
which, using Eq.\ (\ref{eq:exp-geod-dist}), gives
\begin{equation}\label{HS}
H^S(x)=\frac{\mu}{\epsilon_0} - \frac{\mu S_1}{2\epsilon_0^3} + O(\delta x).
\end{equation}

Let us now introduce the field $H^{\rm full}(x)\equiv\frac{1}{2}h_{\alpha\beta}^{\rm full} v^\alpha v^\beta$. Then, noting $v^{\alpha}(x\to z)=u^{\alpha}$, we obtain using Eqs.\ (\ref{HRdef}) and (\ref{HR})
\begin{equation}\label{modesum1}
H^R=\lim_{x\to z}\left[H^{\rm full}(x)-H^S(x)\right].
\end{equation}
As in the standard mode-sum regularization prescription, we formally expand both $H^{\rm full}(x)$ and $H^S(x)$ in spherical harmonics (on a 2-sphere of constant $r,t$), and write Eq.\ (\ref{modesum1}) in the form 
\begin{equation}\label{modesum2}
H^R=\lim_{x\to z}\sum_{l=0}^{\infty}\left[H_l^{\rm full}(x)-H^S_l(x)\right],
\end{equation}
where $H_l^{{\rm full}/S}$ represent the total $l$-mode contribution (summed over $m$) to $H^{{\rm full}/S}$.
It can be shown, based on the form of the local expansion in Eq.\ (\ref{HS}), that each of the $l$-modes $H_l^{S}(x)$ is continuous (though not differentiable) at $x\to z$; and that, at large $l$, $H_l^{S}(z)$ admits an expansion of the form 
\begin{equation}\label{HlS}
H_l^{S}(z)= B_H + C_H/l +O(l^{-2}),
\end{equation}
where the coefficients $B_H$ and $C_H$ depend on $z$ but not on $l$. Since $H^{\rm full}(x)-H^S(x)$ is a smooth function, the multipole sum in Eq.\ (\ref{modesum2}) must converge uniformly and faster than any power-law in $1/l$, and it follows that $H_l^{\rm full}(z)$, too, must admit the large-$l$ expansion 
\begin{equation}
H_l^{\rm full}(z)= B_H + C_H/l +O(l^{-2}),
\end{equation}
with the same $B_H$ and $C_H$. This allows us to reexpress Eq.\ (\ref{modesum2}) as a sum of two convergent series, using the familiar mode-sum form 
\begin{equation}\label{modesum3}
H^R=\sum_{l=0}^{\infty}\left[H_l^{\rm full}(z)-B_H - C_H/l\right]-D_H,
\end{equation}
with 
\begin{equation}\label{DH}
D_H=\sum_{l=0}^{\infty}\left[H_l^{\rm S}(z)-B_H - C_H/l\right].
\end{equation}
The parameters $B_H$, $C_H$ and $D_H$ are akin to the standard self-force ``regularization parameters''---which we have attempted to reflect in our notation. The absence of a regularization term $\propto l$ is related to the fact that $H^R$, unlike the self force, does not involve derivatives of the metric perturbation.

The derivation of the parameters $B_H$, $C_H$ and $D_H$ follows closely the method detailed in Secs.\ VII and VIII of Ref.\ \cite{Barack:2001gx}. It is based on expanding $H^S(x)$ [given in Eq.\ (\ref{HS})] in spherical harmonics, and then evaluating the $l$-mode contribution at the limit $x\to z$. First, one notes that the terms of $O(\delta x)$ and higher in Eq.\ (\ref{HS}) can be discarded as they cannot affect the value of $H^R$ in Eq.\ (\ref{modesum1}). [The individual $l$-modes of the $O(\delta x)$ terms may well be nonzero, but their summed contribution must vanish at the limit $x\to z$.] We can therefore write 
\begin{equation}\label{modesum1approx}
H^R=\lim_{x\to z}\left[H^{\rm full}(x)-H^{S,{\rm approx}}(x)\right],
\end{equation}
where
\begin{equation}\label{HSapprox}
H^{S,{\rm approx}}(x)\equiv \frac{\mu}{\epsilon_0} - \frac{\mu S_1}{2\epsilon_0^3},
\end{equation}
and consider the $l$-mode contributions from $H^{S,{\rm approx}}$ instead of those from $H^{S}$.
Considering first the term $\propto S_1$, one readily shows based on a simple symmetry consideration (cf.\ Sec.\ VII.B of \cite{Barack:2001gx}) that it yields no contribution to $H_l^{S,{\rm approx}}(z)$. The validity of this consideration does not depend on the explicit form of $S_1$ (but only on the fact that it is cubic in $\delta x$), and it is the same consideration that leads one to conclude the vanishing of the `$C$' parameter in the self-force case. 

Hence, the sole contribution to $H_l^{\rm S,{\rm approx}}(z)$ comes from the term $\mu/\epsilon_0$ in Eq.\ (\ref{HSapprox}). This contribution is easily evaluated in explicit form using the method of Sec.\ VII.C of \cite{Barack:2001gx}, and one finds, crucially, that it is $l$-independent. One therefore identifies this contribution with the parameter $B_H$
[recall Eq.\ (\ref{HlS})], and moreover concludes that $C_H=0$. Furthermore, since the above implies $H_l^{\rm S,{\rm approx}}(z)-B_H-C_H/l=0$ for all $l$, it follows from the definition in Eq.\ (\ref{DH}) that $D_H=0$. Explicitly, one finds, in summary,
\begin{equation}
B_H =
\frac{2\mu}{\pi\sqrt{r_0^2+{L}^2}}\,
{\rm ellipK}\left(\frac{{L}^2}{r_0^2+{L}^2}\right), \quad\quad
C_H=D_H=0,
\end{equation}
where, recall, ${\rm ellipK}(\cdot)$ is the complete elliptic integral of the first kind.
Equation (\ref{modesum3}) thus reduces to 
\begin{equation}\label{modesum4}
H^R=\sum_{l=0}^{\infty}\left[H_l^{\rm full}(z)-B_H\right].
\end{equation}

To calculate $H^R$ along the orbit we begin by recording the values of the (Lorenz-gauge) tensor harmonic $l$-modes $h_{\alpha\beta}^{{\rm full},l}$ generated by our time-domain code (the $l$-modes are continuous at the orbit, so these values are well defined). We then construct the full $l$-mode contributions $H^{\rm full}_l(z)$ (as functions along the orbit) using
\begin{equation}
H^{\rm full}_l(z)\equiv\frac{1}{2}h_{\alpha\beta}^{{\rm full},l} u^\alpha u^\beta.
\end{equation}
We use these modes as input in the mode-sum formula (\ref{modesum4}), which yields $H^R$. In practice we compute numerically only the first $20$ or so $l$-modes. We estimate the contribution from the uncomputed large-$l$ tail by fitting a power-law model ($a/l^2+b/l^4$) to the numerical data, then add the estimated tail contribution to the mode-sum. In this we follow the same procedure as for the GSF, and we refer the reader to Ref.\ \cite{Barack:2010tm} for details.

\section{Numerical values for the circular-limit GSF coefficients}\label{App:circ}

\nopagebreak
Table \ref{table:circ} gives numerical values for the GSF coefficients $F_\circ^r$, $F_1^r$ and $F^1_\varphi$ appearing in Eqs.\ (\ref{freqratio}) and (\ref{hatOmegaf}), for a sample of circular-orbit radii $r_\circ$. How these values are extracted in practice from the numerical GSF data is explained in Ref.\ \cite{Barack:2010tm}. We give these values here for the benefit of readers who wish to reproduce the GSF curves shown in Fig.\ \ref{fig:NR}, or wish to obtain similar curves for other values of the mass ratio $q$.  

\begin{table}[htb]

\begin{tabular}{llll}
\hline\hline
$r_{\circ}/M$ & $q^{-2}F_\circ^r\times 10^3$
 & $q^{-2}F_1^r\times 10^3$ & $\mu^{-2}F^1_\varphi$   \\
\hline\hline
$80$  & $0.29897884(1)$ & $0.58707883(1)$ & $-0.318667(3)$ \\
$50$  & $0.74494870(4)$ & $1.4474199(1)$  & $-0.391467(4)$ \\
$40$  & $1.1428834(1)$  & $2.2055600(3)$  & $-0.429392(1)$ \\
$30$  & $1.96981742(2)$ & $3.7606729(4)$  & $-0.4805562(2)$ \\
$25$  & $2.76576978(2)$ & $5.2382103(4)$  & $-0.5136676(2)$ \\
$20$  & $4.1570578(2)$  & $7.7897336(5)$  & $-0.5540440(9)$ \\
$19$  & $4.5584736(5)$  & $8.520716(7)$   & $-0.563208(3)$ \\
$18$  & $5.020136(1)$   & $9.359448(3)$   & $-0.572799(2)$ \\
$17$  & $5.554467(2)$   & $10.328290(6)$  & $-0.582853(5)$ \\
$16$  & $6.177153(2)$ & $11.455829(7)$  & $-0.593422(1)$ \\
$15$  & $6.908156(3)$   & $12.779174(6)$  & $-0.604583(4)$ \\
$14$  & $7.7730719(9)$  & $14.34739(1)$   & $-0.616448(2)$ \\
$13$  & $8.804901(2)$   & $16.22683(1)$   & $-0.629220(3)$ \\
$12$  & $10.046259(3)$  & $18.50998(2)$   & $-0.643288(3)$ \\
$11$  & $11.551775(4)$  & $21.33090(3)$   & $-0.659417(3)$ \\
$10$  & $13.389514(2)$  & $24.89511(5)$   & $-0.679283(4)$ \\
$9.0$ & $15.637164(4)$  & $29.54408(9)$   & $-0.706774(4)$ \\
$8.5$ & $16.936718(5)$  & $32.4544(1)$    & $-0.725949(5)$ \\
$8.0$ & $18.357927(4)$  & $35.9162(2)$    & $-0.751553(2)$ \\
$7.5$ & $19.89068(1)$   & $40.1309(2)$    & $-0.78753(2)$ \\
$7.0$ & $21.49921(1)$   & $45.4273(2)$    & $-0.84103(2)$ \\
$6.5$ & $23.09361(1)$   & $52.3795(3)$    & $-0.92545(3)$ \\
$6.0$ & $24.4665(1)$    & $62.095(1)$     & $-1.0665(8)$ \\
\hline\hline
\end{tabular}
\caption{Numerical values for the circular-limit GSF coefficients appearing in Eqs.\ (\ref{freqratio}) and (\ref{hatOmegaf}). $r_{\circ}$ is the radius of the limiting circular orbit. Parenthetical figures show the estimated uncertainty in the last displayed decimals. Data from this table were used to generate the GSF curves in Fig.\ \ref{fig:NR}. 
}
\label{table:circ}
\end{table}


\end{document}